\documentstyle[amssymb,preprint,aps]{revtex}
\begin{document}
\title{Unified theory of quantum many-particle systems}
\author{Yu-Liang Liu}
\address{Center for Advanced Study, Tsinghua University, Beijing 100084, \\
People's Republic of China}
\maketitle

\begin{abstract}
Using eigen-functional bosonization method, we study quantum many-particle
systems, and show that the quantum many-particle problems end in to solve
the differential equation of the phase fields which represent the particle
correlation strength. Thus, the physical properties of these systems are
completely determined by the differential equation of the phase fields. We
mainly focus on the study of D-dimensional electron gas with/without
transverse gauge fields, two-dimensional electron gas under an external
magnetic field, D-dimensional boson systems, a D-dimensional Heisenberg
model and a one-band Hubbard model on a square lattice, and give their exact
(accurate for Heisenberg model) functional expressions of the ground state
energy and action, and the eigen-functional wave functions of the
fermions/bosons. With them, we can calculate a variety of correlation
functions of the systems, such as single particle Green's functions and
their ground state wave functions. In present theoretical framework, we can
unifiably represent the Landau Fermi liquid, non-Fermi liquid ($D\geq 2$)
and Tomonaga-Luttinger liquid.
\end{abstract}


\begin{center}
\vspace{1cm}

{\bf I. Introduction}

\bigskip
\end{center}

Quantum many-particle systems are the main topics of the condensed matter
physics, in which strongly correlated electron systems are the most
interesting and hard problems, such as heavy fermion systems, high Tc
cuprate superconductors, fractional quantum Hall effects, and some
one-dimensional interacting fermion systems. In general, the quantum
many-particle (fermion) systems can be divided into two categories, one is
represented by the Landau Fermi liquid theory\cite{1,2}, and may be called
as weakly correlated systems, and another one is represented by non-Fermi
liquid theory, such as the Tomonaga-Luttinger liquid theory\cite{3,4,5} and
marginal Fermi liquid theory\cite{6,7}, and may be called as strongly
correlated fermion systems. Thus there explicitly exists a key parameter
hidden in the quantum many-particle systems, which represents the
fermion/boson correlation which is produced by the particle interaction.
However, the traditional perturbation theories, such as Hartree-Fock
approximation, mean field theory, random-phase approximation (RPA) and
renormalization group theory, are starting from the bare Green's function to
perturbatively treat the interaction (potential) terms of the particles by
taking some low order Feynman diagrams, i.e., exactly treating the kinetic
energy term and perturbatively treating the particle interaction. In these
methods, there does not exist a parameter used to represent the particle
correlation which completely determines the physical properties of the
strongly correlated systems. Thus, the previous perturbation methods are
successful in treating the weakly correlated systems, but using them to
treat the strongly correlated systems, one can meet some serious problems,
because in these systems it is difficult to find a suitable small quantity
as a perturbation expansion parameter.

It becomes clear that if one wants to unifiably represent the weakly and
strongly correlated systems, one must find a well-defined parameter which
can represent the particle correlation strength. Due to the particle
correlation is produced by the particle interaction, it is natural that we
are starting from the interaction (potential) terms to exactly or
perturbatively treat the kinetic energy term of the particles, i.e., we
first exactly treat the interaction terms, then with them we exactly or
perturbatively treat the kinetic energy term. In this way, the basic
ingredient of the systems is particle density field rather than particle
field operators, like usual bosonization method in treating one-dimensional
interacting fermion systems. In Ref.\cite{8}, we proposed an
eigen-functional bosonization method, in which there naturally appears a key
parameter called the phase field, and with it we can unifiably represent the
weakly and strongly correlated electron gases, i.e., the Landau Fermi liquid
theory and the non-Fermi liquid theory can be unified under our
eigen-functional bosonization theory.

According to the Hohenberg-Kohn theorem\cite{9}, the ground state energy of
the quantum many-particle systems is uniquely determined by their ground
state particle density. This theorem does not tell us how to construct the
ground state energy by the particle density, but it clearly tells us that
the ground state energy can be exactly represented by the particle density.
However, in the Kohn-Sham scheme\cite{10,11}, we can only approximately
obtain the expression of the ground state energy as a functional of the
ground state particle density, because in this scheme the exchange
correlation energy term is unclear. Only for small exchange correlation
energy can one use the Kohn-Sham scheme to obtain reliable results, i.e.,
one cannot use it to study the strongly correlated systems, where the
exchange correlation energy is large . Thus it is very desirable to write
out an exact expression of the ground state energy of the quantum
many-particle systems as a functional of the particle density. In fact, in
usual bosonization representation of one-dimensional interacting fermion
systems\cite{12,13}, we learn how to construct the Hamiltonian by the
fermion density operators, where we use the fermion density field to
uniquely represent the kinetic energy and the fermion field operators. With
the eigen-functional bosonization theory, we can treat any quantum
many-particle system by using the fermion/boson density field to represent
the kinetic energy and the eigen-functional wave functions of the
fermions/bosons. In Ref.\cite{13'}, we demonstrated how to use the particle
density field to exactly represent the ground state energy of the quantum
many-particle systems.

In this paper, we shall give a detail description of the eigen-functional
bosonization method, and apply it to quantum many-particle systems in
continuous coordinate space and some lattice models, such as the electron
gases with or without transverse gauge fields, the two-dimensional electron
gas under an external magnetic field, the interacting boson systems, the
Heisenberg model and one-band Hubbard model. We clearly show that the
problems of these systems end in to solve the differential equation of the
phase fields which completely controls the physical properties of the
systems, thus the quantum many-particle systems can be unifiably represented
by the eigen-functional bosonization theory. In section II, we give the
exact expressions of the ground state energy as a functional of the particle
density field for different quantum many-particle systems. These expressions
are universal, i.e., they are valid not only for weak fermion/boson
interactions, but also for strong fermion/boson interactions. (1). we give
the exact expression of the ground state energy as a functional of the
electron density field for electron gases with and without transverse gauge
fields, respectively; (2). we give the exact functional expression of the
ground state energy for interacting boson systems. (3). we give the exact
functional expression of the ground state energy of one-band Hubbard model
on a square lattice, and the approximate functional expression of the ground
state energy for the Heisenberg model. In section III, we focus on to
calculate the action of the quantum many-particle systems. (1). we give the
exact functional expression of the action for the electron gas with the
transverse gauge fields, and the eigen-functional wave functions of the
electrons. Under linearization approximation (only keeping the linear terms
in solving the differential equation of the phase fields), we obtain its
effective action which is the same as that obtained by usual random-phase
approximation (RPA). (2). we give the exact functional expression of the
action of the one-band Hubbard model on the square lattice, and the
eigen-functional wave functions of the electrons. In section IV, we
calculate the ground state wave function of the interacting boson systems,
which is very similar to the correlated basis functions used in the study of
the liquid $^4He$. We also approximately calculate the ground state wave
function of the two-dimensional electron gas under the external magnetic
field, which has a little similarity with Laughlin's trial wave-functions%
\cite{14,15}. In section V, by calculating the correlation function of the
phase fields, we show that the imaginary part of the phase fields does
represent the particle correlation strength. We also demonstrate that if the
non-Fermi liquid behavior of the systems is derived from the electron
Coulomb interactions, the correlation function of the phase fields has a
log-type expression; If their non-Fermi liquid behavior is derived from the
transverse gauge fields, the correlation function of the phase fields has a
power-law form, $t^\alpha $, $\alpha <1$. Under our present theoretical
framework, the Landau Fermi liquid, non-Fermi liquid and Tomonaga-Luttinger
liquid can be unifiably represented. We give some discussions and our
conclusion in section VI.

\begin{center}
\bigskip

{\bf II. Exact functional expression of the ground state energy}

\bigskip
\end{center}

In this section, we shall give the exact expression of the ground state
energy as a functional of the particle density field for a variety of
quantum many-particle systems. In general, a quantum many-particle system
can be described by the Hamiltonian, $H_{\text{T}}=H_0+H_{\text{ex}}+H_{%
\text{spin}}$, where $H_0$ presents the kinetic energy and the
density-density interactions of the particles, $H_{\text{ex}}$ presents the
interactions between the system and external fields, and $H_{\text{spin}}$
presents the spin-spin interactions among the particles, which is usually
written out on lattice sites. In the following we consider the systems with
these different particle interactions.

\begin{center}
\smallskip

\smallskip A. Electron gas

\medskip
\end{center}

For simplicity, we first consider an electron gas with the Hamiltonian (here
omitting the spin label), 
\begin{equation}
H_0={\ \psi ^{\dagger }(x)\left( \frac{\hat{p}^2}{2m}-\mu \right) \psi (x)+%
\frac 12\int d^Dyv(x-y)\rho (y)\rho (x)}  \label{1a}
\end{equation}
where $\hat{p}=-i\hbar {\bf \nabla }$, $\rho (x)=\psi ^{\dagger }(x)\psi (x)$
is the electron density operator, and $D$ the dimensions of the system. In
usual perturbation theories, such as Hartree-Fock approximation,
Random-Phase approximation (RPA) and renormalization group theory, one
mainly focuses on how to perturbatively treat the density-density
interaction (four-fermion interaction) term, and how to get more accurate
results by considering some special Feynman diagrams. However, these methods
do not always work for low-dimensional strongly correlated systems.
According to the Hohenberg-Kohn theorem and the bosonization representation
of one-dimensional fermion systems, we know that the ground state energy
and/or the Hamiltonian can be uniquely represented by the ground state
particle density, thus we can also study this system by perturbatively or
exactly treating the kinetic energy term of the electrons. In order to
calculate the ground state energy, we introduce a Lagrangian multiplier (or
Hubbard-Stratonovich field) $\phi (x)$ which takes $\rho (x)=\psi ^{\dagger
}(x)\psi (x)$ as a constraint condition, then the Hamiltonian (\ref{1a}) can
be re-written as, 
\begin{eqnarray}
H_0[\phi ,\rho ] &=&{\ \psi ^{\dagger }(x)\left( \frac{\hat{p}^2}{2m}-\mu
+\phi (x)\right) \psi (x)}  \nonumber \\
&-&{\ \phi (x)\rho (x)+\frac 12\int d^Dyv(x-y)\rho (y)\rho (x)}  \label{2}
\end{eqnarray}
where $\phi (x)$ and $\rho (x)$ are independent boson fields, and the
Hamiltonian $H_0[\phi ,\rho ]$ only has the quadratic term of the electron
field operator $\psi (x)$. It is clear that the kinetic energy and potential
energy of the electrons are completely separated, and the electrons move in
the ''effective potential'' $\phi (x)$. Only in this way (four-fermion
interaction replaced by the ''effective potential''), we can exactly treat
the kinetic energy term. We would like to point out that after introducing
the boson field $\phi (x)$, the Hilbert space of the system is enlarged,
because the electron density field $\rho (x)$ and the electron field
operator now become independent of each other, and some unphysical electron
density field $\rho (x)\neq \psi ^{\dagger }(x)\psi (x)$ also appears in the
Hamiltonian \ref{2}. To only keep the physical electron density field $\rho
(x)$, we must add a constraint condition to the system (see below) $\delta
E[\phi ,\rho ]/\delta \phi (x)=0$. In analogy to the eigen-functional
bosonization theory\cite{8}, here we solve the eigen-functional equation of
the ''propagator'' operator (in correspondence with below $\widehat{M}(x,t)$
which is a real propagator operator) $\widehat{M}(x)=\widehat{p}^2/(2m)-\mu
+\phi (x)$, 
\begin{equation}
\left( \frac{\hat{p}^2}{2m}-\mu +\phi (x)\right) \Psi _k(x,[\phi ])=E_k[\phi
]\Psi _k(x,[\phi ])  \label{3}
\end{equation}
where $\Psi _k(x,[\phi ])$ are the eigen-functional wave functions of the
electrons for a definite boson field $\phi (x)$. Using the Helmann-Feynman
theorem, we have the following expression of the eigen-values, 
\begin{eqnarray}
E_k[\phi ] &=&\varepsilon (k)+\Sigma _k[\phi ]  \nonumber \\
\Sigma _k[\phi ] &=&{\ \int_0^1d\xi \int d^Dx\phi (x)|\Psi _k(x,[\xi \phi
])|^2}  \label{4}
\end{eqnarray}
where $\varepsilon (k)=(\hbar k)^2/(2m)-\mu $. The eigen-functionals $\Psi
_k(x,[\phi ])$ have the following exact expression, 
\begin{equation}
\Psi _k(x,[\xi \phi ])=\frac{A_k}{L^{D/2}}e^{i{\bf k}\cdot {\bf x}%
}e^{Q_k(x,\xi )}  \label{5}
\end{equation}
where $A_k$ is the normalization constant, and the phase fields $Q_k(x,\xi )$
satisfy usual (static) Eikonal equation, 
\begin{equation}
\left( \frac{\hat{p}^2}{2m}+\frac \hbar m{\bf k}\cdot \hat{p}\right)
Q_k(x,\xi )+\frac{[\hat{p}Q_k(x,\xi )]^2}{2m}+\xi \phi (x)=\Sigma _k[\phi ]
\label{6}
\end{equation}
It is worthily noted that the eigen-functionals are composed of two parts,
one represents the free electron, and another one represents the
contributions from other electrons by the interaction potential $v(x-y)$,
therefore it is clear that the eigen-functionals are the eigen-functional
wave functions of the interacting electrons corresponding to the definite
boson field $\phi \left( x\right) $. With them, we can construct the ground
state wave-function, and calculate a variety of correlation functions of the
system by taking a functional average over the boson field $\phi \left(
x\right) $.

The electron field operators $\psi ^{\dagger }(x)$ and $\psi (x)$ can be
represented as, respectively, 
\begin{eqnarray}
\psi ^{\dagger }(x) &=&{\sum_k\Psi _k^{*}(x,[\phi ])\hat{c}_k^{\dagger }} 
\nonumber \\
\psi (x) &=&{\sum_k\Psi _k(x,[\phi ])\hat{c}_k}  \label{7}
\end{eqnarray}
where $\hat{c}_k^{\dagger }$ ($\hat{c}_k$) is the electron's creation (
annihilation) operator with momentum $\hbar k$. With equations. (\ref{3})
and (\ref{7}), we can obtain the exact expression of ground state energy as
a functional of the boson field $\phi (x)$ and the electron density $\rho
(x) $, 
\begin{eqnarray}
E[\phi ,\rho ] &=&{\int d^Dx<H}_0{[\phi ,\rho ]>}  \nonumber \\
&=&{\sum_k\theta (-E_k[\phi ])E_k[\phi ]-\int d^Dx\phi (x)\rho (x)+\frac 12%
\int d^Dxd^Dyv(x-y)\rho (x)\rho (y)}  \label{8}
\end{eqnarray}
where the boson field $\phi (x)$ is determined by the condition $\delta
E[\phi ,\rho ]/\delta \phi (x)=0$, which leads to the equation, 
\begin{equation}
\sum_k\theta (-E_k[\phi ])\left( G_k(x)+\int d^Dy\phi (y)\frac{\delta G_k(y)%
}{\delta \phi (x)}\right) =\rho (x)  \label{9}
\end{equation}
where $G_k(x)=\int_0^1d\xi e^{2Q_k^R(x,\xi )}/\int d^Dxe^{2Q_k^R(x,\xi )}$,
and $Q_k(x,\xi )=Q_k^R(x,\xi )+iQ_k^I(x,\xi )$. This equation shows that the
boson field $\phi (x)$ is the functional of the electron density field $\rho
(x)$. The chemical potential is determined by the constraint equation, 
\begin{equation}
\sum_k\theta (-E_k[\phi ])=N  \label{10}
\end{equation}
where $N$ is the total electron number. With equation (\ref{9}), the ground
state energy can be represented as another form, in which it is only the
functional of the electron density field ($\phi (x)$ is the functional of $%
\rho (x)$ determined by (\ref{9})), 
\begin{equation}
E_g[\rho ]=E_0+\frac 12\int d^Dxd^Dy\left( v(x-y)\rho (x)\rho (y)-2\phi
(x)\phi (y)\sum_k\theta (-E_k[\phi ])\frac{\delta G_k(y)}{\delta \phi (x)}%
\right)  \label{11}
\end{equation}
where $E_0=\sum_k\theta (-E_k[\phi ])\varepsilon (k)$, and the self-energy
can be written as a simple form, $\Sigma _k[\phi ]=\int d^Dx\phi (x)G_k(x)$.
The equations (\ref{6}), (\ref{9}), (\ref{10}) and (\ref{11}) give the exact
expression of the ground state energy as a functional of the electron
density, we can numerically calculate it by taking $\delta E_g[\rho ]/\delta
\rho (x)=0$. These equations have more advantages than that in the Kohn-Sham
scheme, because that, 1). the expression of the ground state energy as the
functional of the electron density is exact, with it we can exactly
determine the ground state energy and the ground state electron density; 2).
with these equations, we can easily estimate the contributions from high
order terms, and obtain enough accurate results we hoped for some special
considerations; 3). Using these equations to calculate the ground state
energy and the ground state electron density, we may need much less computer
time than that in the Kohn-Sham scheme, because here we do not need to
self-consistently solve the eigen-equation of single electron. The ground
state wave-function of the system can be obtained by the eigen-functionals,
however, we cannot write it as a simple form, because the phase field $%
Q_k(x,\xi =1)$ is the function of the momentum $k$.

The expression of the ground state energy (\ref{11}) is valid not only for
Landau Fermi liquid (weak correlation systems), but also for non-Fermi
liquid (strongly correlated systems), because it is universal for weak and
strong electron interactions. The phase field $Q_k(x,\xi )$ is a key
parameter for unifiably representing the Landau Fermi liquid and non-Fermi
liquid. Its imaginary part represents the electron correlation strength, and
its real part ($D\geq 2$) only contributes to the ground state energy and
the action of the systems, this can be clearly seen in the expression of the
ground state energy (\ref{11}).

\begin{center}
\smallskip

\smallskip B. Electron motion under an external magnetic field

\smallskip
\end{center}

We now consider the electron motion under an external magnetic field ${\bf B}%
=(0,0,B)$, where the response of the system to the magnetic field cannot be
written as a simple density form. It is well-known that for enough strong
magnetic field, in low temperature limit a two-dimensional electron gas
shows the fractional quantum Hall effects due to the Coulomb interaction of
the electrons. Here we only give the exact expression of the ground state
energy of this system, and do not compare it with that obtained by
Laughlin's trial wave-functions\cite{14}, because if so it needs more detail
numerical calculations. Under the magnetic field, the Hamiltonian (\ref{2})
becomes, 
\begin{eqnarray}
H[\phi ,\rho ] &=&{\psi ^{\dagger }(x)\left( \frac 1{2m}(\hat{p}+\frac ec%
{\bf A})^2-\mu +\phi (x)\right) \psi (x)}  \nonumber \\
&-&{\phi (x)\rho (x)+\frac 12\int d^2yv(x-y)\rho (y)\rho (x)}  \label{12}
\end{eqnarray}
where the gauge field ${\bf A}=(-yB/2,xB/2,0)$, and for simplicity we take $%
D=2$. The eigen-functional equation corresponding to equation (\ref{3})
reads, 
\begin{equation}
\left( \frac 1{2m}(\hat{p}+\frac ec{\bf A})^2-\mu +\phi (x)\right) \Psi
_{nl}(x,[\phi ])=E_{nl}[\phi ]\Psi _{nl}(x,[\phi ])  \label{13}
\end{equation}
and the eigen-values are 
\begin{eqnarray}
E_{nl}[\phi ] &=&\epsilon _n+\Sigma _{nl}[\phi ]  \nonumber \\
\Sigma _{nl}[\phi ] &=&\int_0^1{d\xi }\int {d^2x\phi (x)|\Psi _{nl}(x,[\xi
\phi ])|^2}  \label{14}
\end{eqnarray}
where $\epsilon _n=\hbar \omega _0(n+1/2)-\mu $, $n=0,1,2,...$, $\omega
_0=eB/(mc)$ is the cyclotron frequency, and $\Psi _{nl}(x,[\phi ])$ are the
eigen-wave functions of the electrons under the external magnetic field $B$
for a definite boson field $\phi (x)$. It is well-known that at $\phi (x)=0$
the eigen-equation (\ref{13}) has exact solutions, 
\begin{equation}
\Psi _{nl}(x,[0])=\psi _{nl}(x)=a_{nl}\left( z-\frac \partial {\partial z^{*}%
}\right) ^l\left( z^{*}-\frac \partial {\partial z}\right) ^ne^{-zz^{*}}
\label{14a}
\end{equation}
where $l=0,1,2,..,L_{max}=\Phi /\Phi _0$, $\Phi =BS$ is the total flux, $%
\Phi _0=2\pi \hbar c/e$ is the flux quantum, $z=(x-iy)/(2l_B)$, $%
z^{*}=(x+iy)/(2l_B)$, $l_B=(\hbar c/(eB))^{1/2}$ is the magnetic length, and 
$a_{nl}$ a normalization constant. Therefore, the eigen-functionals $\Psi
_{nl}(x,[\phi ])$ have the following exact form, 
\begin{equation}
\Psi _{nl}(x,[\xi \phi ])=A_{nl}\psi _{nl}(x)e^{Q_{nl}(x,\xi )}  \label{15}
\end{equation}
where $A_{nl}$ is a normalization constant, and the phase field $%
Q_{nl}(x,\xi )$ satisfy (static) Eikonal-type equation, 
\begin{equation}
\left( \frac{\hat{p}^2}{2m}+(\frac e{mc}{\bf A}+{\bf a}_{nl})\cdot \hat{p}%
\right) Q_{nl}(x,\xi )+\frac{[\hat{p}Q_{nl}(x,\xi )]^2}{2m}+\xi \phi
(x)=\Sigma _{nl}[\phi ]  \label{16}
\end{equation}
where ${\bf a}_{nl}=(1/m)\hat{p}\ln \psi _{nl}(x)$.

Following the above same procedures, we have the exact expression of the
ground state energy as the electron density field $\rho (x)$, 
\begin{equation}
E_g[\rho ]=E_0+\frac 12\int d^2xd^2y\left( v(x-y)\rho (x)\rho (y)-2\phi
(x)\phi (y)\sum_{nl}\theta (-E_{nl}[\phi ])\frac{\delta G_{nl}(y)}{\delta
\phi (x)}\right)  \label{17}
\end{equation}
where $E_0=\sum_{nl}\theta (-E_{nl}[\phi ])\epsilon _n$, and the boson field 
$\phi (x)$ is the functional of the electron density $\rho (x)$, and
determined by the equation, 
\begin{equation}
\sum_{nl}\theta (-E_{nl}[\phi ])\left( G_{nl}(x)+\int d^2y\phi (y)\frac{%
\delta G_{nl}(y)}{\delta \phi (x)}\right) =\rho (x)  \label{18}
\end{equation}
where $G_{nl}(x)=\int_0^1d\xi |\psi _{nl}(x)|^2e^{2Q_{nl}^R(x,\xi )}/\int
d^2x|\psi _{nl}(x)|^2e^{2Q_{nl}^R(x,\xi )}$, and the self-energy can written
as a simple form, $\Sigma _{nl}[\phi ]=\int d^2x\phi (x)G_{nl}(x)$. The
chemical potential is determined by the constraint condition, 
\begin{equation}
\sum_{nl}\theta (-E_{nl}[\phi ])=N  \label{19}
\end{equation}
where $N$ is the total electron number. The equations (\ref{16}), (\ref{17}%
), (\ref{18}) and (\ref{19}) can be used to exactly determine the ground
state energy and the ground state electron density of the system by taking $%
\delta E_g[\rho ]/\delta \rho (x)=0$. However, the expression of the ground
state energy (\ref{17}) has a great advantages comparing with the Laughlin's
trial wave-functions: 1). it is a microscopic theory expression. 2). it
shows that the odd- and even-denominator's fractional quantum Hall states
have the same expression of the ground state energy. 3). it can exactly
determine the ground state electron density. 4). it is very simple to study
the fractional quantum Hall effects taken place in higher Landau levels ($%
n\geq 1$).

\begin{center}
\smallskip

C. Electron gas with transverse gauge interaction

\smallskip
\end{center}

\smallskip We consider a two-dimensional electron gas represented by the
Hamiltonian with transverse gauge fields ${\bf A}(x)$ (${\bf \nabla }\cdot 
{\bf A}(x)=0$), 
\begin{equation}
H={\psi ^{\dagger }(x)}\left[ {\frac 1{2m}(}\widehat{p}{-g{\bf A})^2-\mu }%
\right] {\psi (x)+}\frac 12{\int }d{^2yv(x-y)\rho (x)\rho (y)}  \label{c1}
\end{equation}
It is well-known that this system has the following low energy behavior, (1)
as $v(x-y)=0$, it shows the non-Fermi liquid behavior; (2) as ${\bf A}(x)=0$
and $v(x)\sim \ln |{\bf x}|$, it also shows the non-Fermi liquid behavior.
Thus the system belongs to the non-Fermi liquid. Introducing the Lagrangian
multiplier $\phi (x)$, we can re-write it as, 
\begin{eqnarray}
H[\phi ,\rho ] &=&{\psi ^{\dagger }(x)}\left[ {\frac 1{2m}(}\widehat{p}{-g%
{\bf A})^2-\mu +\phi (x)}\right] {\psi (x)}  \nonumber \\
&&-\phi (x)\rho (x){+}\frac 12{\int }d{^2yv(x-y)\rho (x)\rho (y)}  \label{c2}
\end{eqnarray}
Then by solving the eigen-functional equation of the ''propagator'' operator 
$\widehat{M}(x)={(}\widehat{p}{-g{\bf A})^2/(2m)-\mu +\phi (x)}$, 
\begin{equation}
\left( \frac 1{2m}{(}\widehat{p}{-g{\bf A})^2-\mu +\phi (x)}\right) \Psi
_k(x,[\phi ])=E_k[\phi ]\Psi _k(x,[\phi ])  \label{c3}
\end{equation}
we can obtain the exact functional expression of the ground state energy.
The solutions of this equation are that, 
\begin{eqnarray}
&&\Psi _k(x,[\xi \phi ])=\frac{A_k}Le^{i{\bf k}\cdot {\bf x}}e^{\overline{Q}%
_k(x,\xi )}  \nonumber \\
&&E_k[\phi ,A]=\epsilon _k+\Sigma _k[\phi ,A]  \label{c4} \\
&&\Sigma _k[\phi ,A]={\ \int_0^1d\xi \int d^2x\Psi _k^{*}(x,[\xi \phi
])[\phi (x)-\frac g{2m}{\bf A\cdot }{(2}\widehat{p}{-g{\bf A})}]\Psi
_k(x,[\xi \phi ])}  \nonumber
\end{eqnarray}
where $\epsilon _k=(\hbar k)^2/(2m)-\mu $, and $A_k$ is a normalization
constant. The phase fields $\overline{Q}_k(x,\xi )$ satisfy the (static)
Eikonal-type equation, 
\begin{eqnarray}
\left( \frac{\hat{p}^2}{2m}+(\frac \hbar m{\bf k}-\frac{\xi g}m{\bf A}%
(x))\cdot \hat{p}\right) \overline{Q}_k(x,\xi )+ &&\frac{[\hat{p}\overline{Q}%
_k(x,\xi )]^2}{2m}  \nonumber \\
+\frac{\xi (g{\bf A}(x))^2}{2m}-\frac{\hbar \xi g}m{\bf k}\cdot {\bf A}(x)+
&&\xi \phi (x)=\Sigma _k[\phi ,A]  \label{cc5}
\end{eqnarray}
and are the functional of the boson field $\phi (x)$ and the gauge fields $%
{\bf A(}x{\bf )}$. In general, this equation cannot be analytically solved
but a series solution. Using the eigen-functionals $\Psi _k(x,[\phi ])$ to
represent the electron field operators $\psi (x)$ and $\psi ^{\dagger }(x)$,
we can obtain the exact expression of ground state energy as a functional of
the boson field $\phi (x)$, the gauge field ${\bf A}(x)$ and the electron
density $\rho (x)$, 
\begin{eqnarray}
E[\phi ,A,\rho ] &=&{\int d^2x<H[\phi ,\rho ]>}  \nonumber \\
&=&{\sum_k\theta (-E_k[\phi ,A])E_k[\phi ,A]-\int d^2x\phi (x)\rho (x)+\frac %
12\int d^2xd^2yv(x-y)\rho (x)\rho (y)}  \label{cc6}
\end{eqnarray}
where the boson field $\phi (x)$ is determined by the condition $\delta
E[\phi ,A,\rho ]/\delta \phi (x)=0$, which leads to the equation, 
\begin{equation}
\sum_k\theta (-E_k[\phi ,A])\left( \overline{G}_k(x)+\int d^2y[\phi (y)\frac{%
\delta \overline{G}_k(y)}{\delta \phi (x)}-\frac g{2m}\overline{G}_k(y){\bf A%
}(y){\bf \cdot }\frac{\delta \left( \widehat{p}\overline{Q}_k(y)\right) }{%
\delta \phi (x)}]\right) =\rho (x)  \label{cc7}
\end{equation}
where $\overline{G}_k(x)=\int_0^1d\xi e^{2\overline{Q}_k^R(x,\xi )}/\int
d^2xe^{2\overline{Q}_k^R(x,\xi )}$, and $\overline{Q}_k(x,\xi )=\overline{Q}%
_k^R(x,\xi )+i\overline{Q}_k^I(x,\xi )$. The gauge fields ${\bf A}(x)$ are
determined by the conditions $J_i(x)=\delta E[\phi ,A,\rho ]/\delta A_i(x)=0$%
, $i=1,2$, which means that the current induced by the gauge fields is zero.
The ground state energy $E[\phi ,A,\rho ]$ (\ref{cc6}) has the similar
functional expression to that of two-dimensional electron gas without gauge
fields (\ref{8}). However, they have different phase field dependence, this
can be clearly seen from their self-energy expressions\smallskip $\Sigma
_k[\phi ]$ and $\Sigma _k[\phi ,A]$, respectively. The gauge fields produce
the term ${\Psi }_k^{*}(x,[\xi \phi ]){\bf A}(x){{\bf \cdot }\widehat{p}\Psi 
}_k(x,[\xi \phi ])=|{\Psi }_k(x,[\xi \phi ])|^2{\bf A}(x){\bf \cdot (\hbar k+%
}\widehat{p}\overline{Q}_k(x,\xi ))${\ in the }self-energy $\Sigma _k[\phi
,A]$, thus in this case the imaginary part of the phase fields may
contribute to the ground state energy. We would like to point that due to
the phase fields $\overline{Q}_k(x,\xi )$ are complex, to keep the
eigen-values $E_k[\phi ,A]$ being real, we must have an additional
constraint condition on the gauge fields ${\bf A}(x){\bf \cdot }\widehat{p}%
\overline{Q}_k^R(x,\xi )=0$. Thus we have three equations to determine two
independent transverse gauge fields. If these equations are independent, we
only have the solution ${\bf A}(x)=0$, i.e., the pure transverse gauge
fields do not contribute to the ground state energy. If they are not
independent each other, we may have non-zero solution, and the transverse
gauge fields contribute to the ground state energy.

\smallskip

\begin{center}
\smallskip D. spin coupling systems

\smallskip
\end{center}

\smallskip Spin coupling systems are strongly correlated systems and
generally represented by the lattice Hamiltonians. For simplicity, we only
consider a D-dimensional spin-1/2 Heisenberg model, 
\begin{equation}
H=J\sum_{<ij>}\widehat{S}_i\cdot \widehat{S}_j  \label{d1}
\end{equation}
where $<ij>$ indicates the summation over the nearest neighbor sites. This
spin-spin interaction represents the four-fermion/boson interaction, and is
difficult to exactly treat it by usual perturbation theory. Here we only
give a way for treating this kind of systems, which may give more useful and
accurate informations than previous perturbation methods.

In general, the spin operator (spin-1/2) $\widehat{S}_i$ can be represented
by the fermion operators $\widehat{S}_i=\frac 12\widehat{f}_{i\alpha
}^{\dagger }{\bf \sigma }_{\alpha \beta }\widehat{f}_{i\beta }$, and the
Hamiltonian can be re-written as\cite{lee} $-\frac 12\sum_{<ij>,\sigma
,\sigma ^{\prime }}\widehat{f}_{i\sigma }^{\dagger }\widehat{f}_{j\sigma }%
\widehat{f}_{j\sigma ^{\prime }}^{\dagger }\widehat{f}_{i\sigma ^{\prime }}$
or $\sum_{<ij>}(\widehat{f}_{i\uparrow }^{\dagger }\widehat{f}_{j\downarrow
}^{\dagger }-\widehat{f}_{i\downarrow }^{\dagger }\widehat{f}_{j\uparrow
}^{\dagger })(\widehat{f}_{j\downarrow }\widehat{f}_{i\uparrow }-\widehat{f}%
_{j\uparrow }\widehat{f}_{i\downarrow })$, thus we can write the Hamiltonian
(\ref{d1}) as another form, 
\begin{eqnarray}
H &=&-\frac J2\sum_{<ij>}\widehat{\chi }_{ij}^{\dagger }\widehat{\chi }%
_{ij}+J\sum_{<ij>}\widehat{D}_{ij}^{\dagger }\widehat{D}_{ij}  \nonumber \\
&&+J\sum_{<ij>}\widehat{S}_i\cdot \widehat{S}_j+\sum_i{\bf \lambda }_i\cdot (%
\widehat{S}_i-\frac 12\widehat{f}_{i\alpha }^{\dagger }{\bf \sigma }_{\alpha
\beta }\widehat{f}_{i\beta })  \label{d2}
\end{eqnarray}
where $\widehat{\chi }_{ij}^{\dagger }=\widehat{f}_{i\sigma }^{\dagger }%
\widehat{f}_{j\sigma }$, and $\widehat{D}_{ij}=\widehat{f}_{j\downarrow }%
\widehat{f}_{i\uparrow }-\widehat{f}_{j\uparrow }\widehat{f}_{i\downarrow }$%
. The Lagrangian multiplier ${\bf \lambda }_i$ take the equation $\widehat{S}%
_i=\frac 12\widehat{f}_{i\alpha }^{\dagger }{\bf \sigma }_{\alpha \beta }%
\widehat{f}_{i\beta }$ as the constraint condition adding to the Hamiltonian
(\ref{d2}), and the fermion operators $\widehat{f}_{i\sigma }$ and the spin
operator $\widehat{S}_i$ are independent each other. In order to treat the
four-fermion interactions in the Hamiltonian (\ref{d2}), we can further
introduce the Lagrangian multipliers $\phi _{ij}^{\prime }$ and $\Delta
_{ij} $ to decouple these interactions, and have the following form, 
\begin{eqnarray}
H &=&-\frac J2\sum_{<ij>}\widehat{\chi }_{ij}^{\dagger }\widehat{\chi }%
_{ij}+J\sum_{<ij>}\widehat{D}_{ij}^{\dagger }\widehat{D}_{ij}  \nonumber \\
&&+J\sum_{<ij>}\widehat{S}_i\cdot \widehat{S}_j+\sum_i{\bf \lambda }_i\cdot (%
\widehat{S}_i-\frac 12\widehat{f}_{i\alpha }^{\dagger }{\bf \sigma }_{\alpha
\beta }\widehat{f}_{i\beta })  \label{d3} \\
&&+\sum_{<ij>}[\phi _{ij}^{\prime }(\widehat{\chi }_{ij}^{\dagger }-\widehat{%
f}_{i\sigma }^{\dagger }\widehat{f}_{j\sigma })+\text{h.c.}%
]+\sum_{<ij>}[\Delta _{ij}(\widehat{D}_{ij}-\widehat{f}_{j\downarrow }%
\widehat{f}_{i\uparrow }+\widehat{f}_{j\uparrow }\widehat{f}_{i\downarrow })+%
\text{h.c.}]  \nonumber
\end{eqnarray}
which is completely equivalent to the Hamiltonian (\ref{d1}). In fact, if we
integrate out the boson fields $\widehat{\chi }_{ij}$ and $\widehat{D}_{ij}$%
, the Hamiltonian (\ref{d3}) becomes usual one\cite{lee}, and the Lagrangian
multipliers $\phi _{ij}^{\prime }$ and $\Delta _{ij}$ become usual
Hubbard-Stratonovich fields.

Now the Hamiltonian (\ref{d3}) only includes the quadratic terms of the
fermion operators, and in principle, we can exactly calculate the ground
state energy. However, in the lattice sites, it is hard to obtain the
eigen-functionals of the fermions and the ground state energy by present
method, we need to mapping this lattice Hamiltonian into a continuous one.
To this end, we take the following approximations, 
\begin{eqnarray}
<\widehat{\chi }_{ij}> &=&a\chi (\frac{x_i+x_j}2),\quad <\widehat{D}%
_{ij}>=aD(\frac{x_i+x_j}2)  \nonumber \\
\phi _{ij}^{\prime } &=&\phi ^{\prime }(\frac{x_i+x_j}2),\quad \Delta
_{ij}=\Delta (\frac{x_i+x_j}2)  \label{d4} \\
\phi _{ij}^{\prime }\widehat{f}_{i\sigma }^{\dagger }\widehat{f}_{j\sigma }
&=&\phi ^{\prime }(\frac{x_i+x_j}2)\widehat{f}_{i\sigma }^{\dagger }\widehat{%
f}_{j\sigma }\cong \gamma \widehat{f}_{i\sigma }^{\dagger }\widehat{f}%
_{j\sigma }+\phi (\frac{x_i+x_j}2)\widehat{f}_{i\sigma }^{\dagger }\widehat{f%
}_{i\sigma }\quad  \nonumber
\end{eqnarray}
where $a$ is the lattice constant, and $\phi ^{\prime }(\frac{x_i+x_j}2%
)=\gamma +\phi (\frac{x_i+x_j}2)$, and obtain the continuous Hamiltonian, 
\begin{eqnarray}
H &=&\sum_k\varepsilon (k)\widehat{f}_{\sigma k}^{\dagger }\widehat{f}%
_{\sigma k}-\int d^Dx[\phi (x)+\phi ^{*}(x)]\widehat{f}_\sigma ^{\dagger }(x)%
\widehat{f}_\sigma (x)+\frac 12\int d^Dx{\bf \lambda }(x)\cdot \widehat{f}%
_\alpha ^{\dagger }(x){\bf \sigma }_{\alpha \beta }\widehat{f}_\beta (x) 
\nonumber \\
&&+\int d^Dx\{2\Delta (x)\widehat{f}_{\uparrow }(x)\widehat{f}_{\downarrow
}(x)-2\Delta ^{*}(x)\widehat{f}_{\uparrow }^{\dagger }(x)\widehat{f}%
_{\downarrow }^{\dagger }(x)+\lambda (x)(\widehat{f}_\sigma ^{\dagger }(x)%
\widehat{f}_\sigma (x)-1)\}  \label{d5} \\
&&+\int d^Dx[aJ|D(x)|^2-\frac{aJ}2|\chi (x)|^2]+\int d^Dx[\phi ^{\prime
}(x)\chi (x)+\Delta (x)D(x)+\text{h.c.}]+H_S  \nonumber
\end{eqnarray}
where $\varepsilon (k)=-2\gamma \sum_l\cos (ak_l),$ $H_S=J\sum_{<ij>}%
\widehat{S}_i\cdot \widehat{S}_j+\sum_i{\bf \lambda }_i\cdot \widehat{S}_i$, 
$\widehat{f}_{i\sigma }=\sqrt{a}\widehat{f}_\sigma (x_i)$, and $\lambda (x)$
is a Lagrangian multiplier which adds the constraint condition at each
lattice site $\widehat{f}_{i\sigma }^{\dagger }\widehat{f}_{i\sigma }=1$.
The parameter $\gamma $ can be approximately determined by usual mean field
theory as taking $\phi (x)=0$, where the parameter $\Delta (x)=\Delta _0$
can also be determined by its self-consistent equation. In the continuous
limit, the Hamiltonian $H_S$ is reduced into one of usual non-linear $\sigma 
$ model under an ''external magnetic field'' ${\bf \lambda }(x)$.

We can further simplify the Hamiltonian (\ref{d5}) by taking $\varepsilon
(k)=\frac{\left( \hbar {\bf k}\right) ^2}{2m}-\mu $, where $m\sim 1/\gamma $
and $\mu $ is the chemical potential of the system. In general, due to $%
\Delta _0\neq 0$, there appears a gap in the excitation spectrum of the
fermions, the dynamic term of the vector fields ${\bf \lambda }(x)$ produced
by the fermions is very small and can be neglected in the low energy
(temperature) region $k_BT\ll \Delta _0$, therefore, we can simply take $%
{\bf \lambda }(x)=0$. We now briefly consider the influence of this
approximation on the spin field. As $J>0$, the system has a long-range or
short-range antiferromagnetic correlation, thus the spin field operator can
be written as $\widehat{S}_i\cong (-1)^iS{\bf \Omega }(x_i)+a{\bf L}(x_i)$,
where ${\bf \Omega }(x_i)\cdot {\bf L}(x_i)=0$ and $|{\bf \Omega }(x_i)|^2=1$%
. In the continuous limit the coupling term becomes $\sum_i{\bf \lambda }%
_i\cdot \widehat{S}_i\rightarrow \frac S{L^{D/2}}\sum_q{\bf \lambda }(q+\pi
/a)\cdot {\bf \Omega }(q)$, thus only the fermion-hole excitations with
large momentum ($\sim \pi /a$) contribute to this coupling term. However,
due to $\Delta _0\neq 0$, these excitations are very small and can be safely
neglected in the low energy limit. Therefore, the Hamiltonian (\ref{d5}) can
be approximately separated into two parts, $H=H_f+H_S$, where $H_f$ mainly
represents the high energy ($\sim \Delta _0$) behavior, and $H_S$ mainly
represents the low energy ($\ll \Delta _0$) behavior of the system. Here we
only consider the contribution of the Hamiltonian $H_f$ to the ground state
energy, where the Hamiltonian $H_f$ reads, 
\begin{eqnarray}
H &=&\int d^Dx\widehat{f}_\sigma ^{\dagger }(x)\{\frac{\widehat{p}^2}{2m}%
-\mu +\lambda (x)-[\phi (x)+\phi ^{*}(x)]\}\widehat{f}_\sigma (x)  \nonumber
\\
&&+2\int d^Dx[\Delta (x)\widehat{f}_{\uparrow }(x)\widehat{f}_{\downarrow
}(x)-\Delta ^{*}(x)\widehat{f}_{\uparrow }^{\dagger }(x)\widehat{f}%
_{\downarrow }^{\dagger }(x)]  \label{d6} \\
&&+\int d^Dx[\frac 2{aJ}|\phi ^{\prime }(x)|^2-\frac 1{aJ}|\Delta
(x)|^2-\lambda (x)]  \nonumber
\end{eqnarray}
Due to appearing the pairing parameter $\Delta (x)=-aJ<\widehat{f}_{\uparrow
}^{\dagger }(x)\widehat{f}_{\downarrow }^{\dagger }(x)>$, the ''propagator''
operator $\widehat{M}(x)$ of the fermions is a $2\times 2$ matrix, and can
be written as $\widehat{M}(x)=\widehat{M}_0+\Phi (x)$, where 
\begin{eqnarray}
\widehat{M}_0 &=&\left( 
\begin{array}{ll}
\frac{\widehat{p}^2}{2m}-\overline{\mu } & \quad -2\Delta _0 \\ 
-2\Delta _0 & \quad -\frac{\widehat{p}^2}{2m}+\overline{\mu }
\end{array}
\right)  \nonumber \\
\Phi (x) &=&\left( 
\begin{array}{ll}
\overline{\lambda }(x)-[\phi (x)+\phi ^{*}(x)] & \quad -2\overline{\Delta }%
^{*}(x) \\ 
-2\overline{\Delta }(x) & \quad -\overline{\lambda }(x)+[\phi (x)+\phi
^{*}(x)]
\end{array}
\right)  \label{d7}
\end{eqnarray}
where $\overline{\mu }=\mu -\lambda _0$, $\overline{\Delta }(x)=\Delta
(x)-\Delta _0$, $\overline{\lambda }(x)=\lambda (x)-\lambda _0$, and the
pairing parameter $\Delta _0$ can be determined by its self-consistent
equation as taking $\phi (x)=\overline{\Delta }(x)=\overline{\lambda }(x)=0$%
. The eigen-functional equation of the ''propagator'' operator $\widehat{M}%
(x)$ reads, 
\begin{equation}
\widehat{M}(x)\Psi _k^{(i)}(x,[\Phi ])=E_k^{(i)}[\Phi ]\Psi _k^{(i)}(x,[\Phi
])  \label{d8}
\end{equation}
where the eigen-values $E_k^{(\pm )}[\Phi ]=\pm E_k+\Sigma _k^{(\pm )}[\Phi
] $, $E_k=\sqrt{\varepsilon ^2(k)+4\Delta _0^2}$, and the self-energy $%
\Sigma _k^{(\pm )}[\Phi ]=\int_0^1d\xi \int d^Dx\Psi _k^{(\pm )*}(x,[\xi
\Phi ])\Phi (x)\Psi _k^{(\pm )}(x,[\xi \Phi ])$. The eigen-functionals $\Psi
_k^{(\pm )}(x,[\xi \Phi ])$ can be generally written as, 
\begin{eqnarray}
\Psi _k^{(+)}(x,[\xi \Phi ]) &=&\frac{A_k^{(+)}}{L^{D/2}}\left( 
\begin{array}{l}
\begin{array}{l}
u_ke^{Q_k^{(+)}(x,\xi )} \\ 
-v_ke^{\overline{Q}_k^{(+)}(x,\xi )}
\end{array}
\end{array}
\right) e^{i{\bf k}\cdot {\bf x}}  \nonumber \\
\Psi _k^{(-)}(x,[\xi \Phi ]) &=&\frac{A_k^{(-)}}{L^{D/2}}\left( 
\begin{array}{l}
\begin{array}{l}
v_ke^{Q_k^{(-)}(x,\xi )} \\ 
u_ke^{\overline{Q}_k^{(-)}(x,\xi )}
\end{array}
\end{array}
\right) e^{i{\bf k}\cdot {\bf x}}  \label{d9}
\end{eqnarray}
where $u_k=(1/\sqrt{2})(1+\varepsilon (k)/E_k)^{1/2}$, $v_k=(1/\sqrt{2}%
)(1-\varepsilon (k)/E_k)^{1/2}$, and $A_k^{(\pm )}$ are the normalization
constants. To calculate the ground state energy, we only consider the
eigen-functionals $\Psi _k^{(-)}(x,[\xi \Phi ])$ which represent the states
occupied by the fermions at zero temperature, and have the differential
equations of the phase fields $Q_k^{(-)}(x,\xi )$ and $\overline{Q}%
_k^{(-)}(x,\xi )$, 
\begin{eqnarray}
\left( \frac{\hat{p}^2}{2m}+\frac{\hbar {\bf k}\cdot \hat{p}}m\right)
Q_k^{(-)}+\frac{[\hat{p}Q_k^{(-)}]^2}{2m}+\xi \Phi (x)+\frac{2u_k}{v_k}%
(\Delta _0-\Delta ^{*}(x)e^{\overline{Q}_k^{(-)}-Q_k^{(-)}}) &=&\Sigma
_k^{(-)}[\Phi ]  \nonumber \\
\left( -\frac{\hat{p}^2}{2m}-\frac{\hbar {\bf k}\cdot \hat{p}}m\right) 
\overline{Q}_k^{(-)}-\frac{[\hat{p}\overline{Q}_k^{(-)}]^2}{2m}-\xi \Phi (x)+%
\frac{2v_k}{u_k}(\Delta _0-\Delta (x)e^{Q_k^{(-)}-\overline{Q}_k^{(-)}})
&=&\Sigma _k^{(-)}[\Phi ]  \label{d10}
\end{eqnarray}
where the pairing parameter $\Delta (x)$ is determined by its
self-consistent equation $\Delta (x)=-aJ<\widehat{f}_{\uparrow }^{\dagger
}(x)\widehat{f}_{\downarrow }^{\dagger }(x)>=\frac{aJ}{2L^D}\sum_k\theta
(-E_k^{(-)}[\Phi ])\frac{\Delta _0|A_k|^2}{E_k}e^{\overline{Q}_k^{(-)}(x,\xi
)+Q_k^{(-)*}(x,\xi )}$. The functional expression of the ground state energy
can be written as a simple form, 
\begin{equation}
E[\lambda ,\phi ,\Delta ]=\sum_k\theta (-E_k^{(-)}[\Phi ])E_k^{(-)}[\Phi
]+\int d^Dx[\frac 2{aJ}|\phi ^{\prime }(x)|^2-\frac 1{aJ}|\Delta
(x)|^2-\lambda (x)]  \label{d11}
\end{equation}
If we can exactly solve the differential equations of the phase fields (\ref
{d10}), we can exactly calculate the self-energy $\Sigma _k^{(-)}[\Phi ]$ by
its definition, then we obtain the ground state energy contributed by the
fermions by taking the constraint conditions, $\delta $ $E[\lambda ,\phi
,\Delta ]/\delta \lambda (x)=\delta $ $E[\lambda ,\phi ,\Delta ]/\delta \phi
(x)=\delta $ $E[\lambda ,\phi ,\Delta ]/\delta \Delta (x)=0$.

\smallskip

\begin{center}
E. One-band Hubbard model

\smallskip
\end{center}

We now consider the one-band Hubbard model on a square lattice, and give the
exact expression of the ground state energy, which is very similar to that
of the two-dimensional electron gas in the continuous coordinates. The
Hamiltonian of the Hubbard model on the square lattice reads, 
\begin{equation}
H=-t\sum_{<ij>,\sigma }(\widehat{c}_{i\sigma }^{\dagger }\widehat{c}%
_{j\sigma }+\widehat{c}_{j\sigma }^{\dagger }\widehat{c}_{i\sigma
})-\sum_{i\sigma }\mu \widehat{c}_{i\sigma }^{\dagger }\widehat{c}_{i\sigma
}+U\sum_in_{i\uparrow }n_{i\downarrow }  \label{e1}
\end{equation}
where $<ij>$ indicates the summation over the nearest neighbor sites, and $U$
is the on-site Coulomb interaction. To decouple the four-fermion
interaction, we introduce the Lagrangian multipliers $\phi _{i\sigma }$
which take $n_{i\sigma }=\widehat{c}_{i\sigma }^{\dagger }\widehat{c}%
_{i\sigma }$ as the constraints to the system, and re-write the Hamiltonian (%
\ref{e1}) as, 
\begin{equation}
H[\phi ,n]=-t\sum_{<ij>,\sigma }(\widehat{c}_{i\sigma }^{\dagger }\widehat{c}%
_{j\sigma }+\widehat{c}_{j\sigma }^{\dagger }\widehat{c}_{i\sigma
})-\sum_{i\sigma }\mu \widehat{c}_{i\sigma }^{\dagger }\widehat{c}_{i\sigma
}-\sum_{i\sigma }\phi _{i\sigma }(n_{i\sigma }-\widehat{c}_{i\sigma
}^{\dagger }\widehat{c}_{i\sigma })+U\sum_in_{i\uparrow }n_{i\downarrow }
\label{e2}
\end{equation}
On the lattice site coordinates, the ''propagator'' operator of the
electrons now can be written as a $N\times N$ matrix, 
\begin{equation}
\left( \widehat{M}_{ij\sigma }\right) =\left( \widehat{M}_{ij}^{(0)}+\phi
_{i\sigma }\delta _{ij}\right)   \label{e3}
\end{equation}
where $\widehat{M}_{ij}^{(0)}=-t(\gamma _{ij}+\gamma _{ji})-\mu \delta _{ij}$%
, $\gamma _{ij}=1$ for $j=i\pm 1$, and $\gamma _{ij}=0$ for other $j$. The
eigen-functional equation of $\widehat{M}_\sigma $ reads, 
\begin{equation}
\widehat{M}_\sigma \Psi _{k\sigma }[\phi ]=E_{k\sigma }[\phi ]\Psi _{k\sigma
}[\phi ]  \label{e4}
\end{equation}
where the eigen-values $E_{k\sigma }[\phi ]=\varepsilon _k+\Sigma _{k\sigma
}[\phi ]$, $\varepsilon _k=-2t(\cos (ak_x)+\cos (ak_y))-\mu ,$ the
self-energy $\Sigma _{k\sigma }[\phi ]=\int_0^1d\xi \Psi _{k\sigma
}^{\dagger }[\xi \phi ]\phi _\sigma \Psi _{k\sigma }[\xi \phi ]$, and $\phi
_\sigma =\left( \phi _{i\sigma }\delta _{ij}\right) $ is a $N\times N$
matrix, in which only the diagonal terms are non-zero. In general, the
eigen-functionals $\Psi _{k\sigma }[\phi ]$ can be written as a simple form, 
\begin{equation}
\Psi _{k\sigma }[\xi \phi ]=\frac{A_k}L\left( 
\begin{array}{l}
e^{i{\bf k}\cdot {\bf x}_1}e^{Q_{k\sigma }(x_{1,}\xi )} \\ 
\quad \quad \ \ \vdots  \\ 
e^{i{\bf k}\cdot {\bf x}_N}e^{Q_{k\sigma }(x_{N,}\xi )}
\end{array}
\right)   \label{e5}
\end{equation}
where $A_k$ is the normalization constant. With the equation (\ref{e4}), the
phase fields $Q_{k\sigma }(x_i,\xi )$ satisfy the following equation, 
\begin{equation}
\gamma _k(1-e^{-Q_{k\sigma }(x_i,\xi )+Q_{k\sigma }(x_{i+1},\xi )})+\gamma
_k^{*}(1-e^{-Q_{k\sigma }(x_i,\xi )+Q_{k\sigma }(x_{i-1},\xi )})+\xi \phi
_{i\sigma }=\Sigma _{k\sigma }[\phi ]  \label{e6}
\end{equation}
where $\gamma _k=t(e^{iak_x}+e^{iak_y})$, and the self-energy $\Sigma
_{k\sigma }[\phi ]$ is self-consistently determined by its definition. The
electron operators $\widehat{c}_{i\sigma }$ can be represented by the
eigen-functionals $\Psi _{k\sigma }[\phi ]$ which are the eigen-functional
wave functions of the electrons for a definite boson field $\phi _{i\sigma }$%
, 
\begin{equation}
\widehat{c}_{i\sigma }=\frac 1L\sum_kA_ke^{i{\bf k}\cdot {\bf x}%
_i}e^{Q_{k\sigma }(x_{i,}\xi )}\widehat{c}_{k\sigma }  \label{e7}
\end{equation}
then we obtain the exact expression of the ground state energy as a
functional of the boson fields $\phi _{i\sigma }$ and the electron density
fields , 
\begin{eqnarray}
E[\phi ,n] &=&<H[\phi ,n]>  \nonumber \\
&=&\sum_{k\sigma }\theta (-E_{k\sigma }[\phi ])E_{k\sigma }[\phi
]-\sum_{i\sigma }\phi _{i\sigma }n_{i\sigma }+U\sum_in_{i\uparrow
}n_{i\downarrow }  \label{e8}
\end{eqnarray}
where the chemical potential of the system is determined by the constraint
condition, 
\begin{equation}
\sum_{k\sigma }\theta (-E_{k\sigma }[\phi ])=N_e  \label{e9}
\end{equation}
where $N_e$ is the total electron number. By taking $\delta E[\phi
,n]/\delta \phi _{i\sigma }=\delta E[\phi ,n]/\delta n_{i\sigma }=0$, we can
exactly calculate the ground state energy and the ground state electron
density of the one-band Hubbard model by using the equations (\ref{e6}), (%
\ref{e8}) and (\ref{e9}). We hope that our present results can provide a
powerful and exact method to study the strongly correlated electron systems
represented by the Hubbard model, and can give some important criterions by
numerical calculations for analytical calculations.

\smallskip

\begin{center}
\smallskip F. Boson systems

\smallskip
\end{center}

For a boson system, such as the liquid $^4He$, at zero temperature it has
Bose-Einstein condensation, and the bosons only occupy the state of the
momentum $k=0$, thus the equation (\ref{10}) is trivial, and the ground
state energy can be written as, 
\begin{equation}
E_g[\rho ]=\frac 12\int d^Dxd^Dy[v(x-y)\rho (x)\rho (y)-2\delta ({\bf x-y}%
)\phi (x)\rho (x)]  \label{f1}
\end{equation}
where we have taken $\mu =\Sigma _0[\phi ]$. The equation (\ref{9}) reduces
into, 
\begin{equation}
G_0(x)+\int d^Dy\phi (y)\frac{\delta G_0(y)}{\delta \phi (x)}=\frac 1N\rho
(x)  \label{f2}
\end{equation}
It is noted that the last term in (\ref{f1}) is the contributions of kinetic
energy of the bosons, and in general it may be non-zero for interacting
boson systems. We can also give the exact functional expression of the
ground state energy of some trapped boson systems, where the procedures are
very similar that we study the two-dimensional electron gas under the
external magnetic field: (1). we first solve the eigen-functional equation
of the ''propagator'' operator $\widehat{M}(x)$ of the bosons under some
external field and/or constraints, and obtain its eigen-values and the
differential equations of the phase fields; (2). using the eigen-functionals
to represent the boson field operators, we obtain the exact expression of
the ground state energy as a functional of the Lagrangian multiplier boson
field $\phi (x)$ and the boson density field $\rho (x)$.

\begin{center}
\bigskip

{\bf III. Exact functional expression of the action}

\bigskip
\end{center}

In this section, we mainly focus on to study two kinds of systems, one is a
D-dimensional electron gas with transverse gauge fields, and another one is
the one-band Hubbard model on a square lattice. In these two cases, we shall
give the exact functional expression of the action and the eigen-functional
wave functions of the electrons. Under linearization approximation (only
keeping the linear terms in solving the differential equation of the phase
fields), we give the effective action which is equivalent to that obtained
by usual RPA method. With them, we can calculate the excitation spectrum and
the variety of correlation functions of the systems, and study their low
energy behavior.

\begin{center}
\bigskip

\smallskip A. Electron gas with transverse gauge field

\bigskip
\end{center}

In general, we consider the system with the Hamiltonian (omitting spin label 
$\sigma $ of the electron operators $c_k$) 
\begin{equation}
H_0={\ \psi ^{\dagger }(x)\left( \frac 1{2m}(\hat{p}-g{\bf A)}^2{\bf -}\mu
\right) \psi (x)+\frac 12\int d^Dyv(x-y)\rho (y)\rho (x)}  \label{3-1}
\end{equation}
where ${\bf A}(x)$ $({\bf \nabla \cdot A}(x)=0)$ are the transverse gauge
fields. The last term represents usual electron's Coulomb interaction
(four-fermion interaction). Introducing the Hubbard-Stratonovich (HS) field $%
\phi (x,t)$ to decouple this four-fermion interaction term, and using the
standard path integral method\cite{19}, we have the action of the system, 
\begin{eqnarray}
S[\phi ,A] &=&{\int dt\int d^dx\left\{ \Psi ^{\dagger }(x,t)\left[ i\hbar 
\frac \partial {\partial t}+\mu -\frac 1{2m}(\hat{p}-g{\bf A)}^2+\phi
(x,t)\right] \Psi (x,t)\right\} }  \nonumber \\
&+&{\frac 1{2TL^D}\sum_{q,\Omega }\frac 1{v(q)}\phi (-q,-\Omega )\phi
(q,\Omega )}  \label{3-2}
\end{eqnarray}
where $\Psi (x,t)$ is the electron field, and only quadratic term of $\Psi
(x,t)$ appears in the action. Therefore, the integration of the electron
field becomes a standard Gaussian form. It is worthily noted that after
introducing the boson field $\phi (x,t)$, the Hilbert space of the system is
enlarged, and the unphysical electron density field $\rho (x,t)\neq \Psi
^{\dagger }(x,t)\Psi (x,t)$ appears in the action \ref{3-2}. To erase the
unphysical electron density field, we must take the functional average over
the boson field $\phi (x,t)$ in calculating a variety of correlation
functions and response functions of the system to external fields.

After integrating out $\Psi (x,t)$, the contribution of ``kinetic'' energy
of the electrons to the action is $-iTr\ln (\widehat{M})$, where $\widehat{M}%
(x,t)=i\hbar \partial _t+\mu -(\hat{p}-g{\bf A)}^2/(2m)+\phi (x,t)$ is the
propagator operator of the electrons. Using the formula, 
\begin{equation}
Tr\ln \left( \widehat{M}\right) =Tr\ln \left( i\hbar \partial _t+\mu -%
\widehat{p}^2/(2m)\right) +\int_0^1d\xi \int dtd^Dx\phi
(x,t)G(x,t;x^{^{\prime }},t^{^{\prime }},[\xi \phi ])|_{\stackrel{t^{\prime
}\rightarrow t}{x^{\prime }\rightarrow x}}  \label{3-3}
\end{equation}
where $\widehat{M}(x,t)G(x,t;x^{^{\prime }},t^{^{\prime }},[\phi ])=\delta (%
{\bf x}-{\bf x}^{\prime })\delta (t-t^{\prime })$, and neglecting the
constant term, we have the following expression of the action, 
\begin{eqnarray}
S[\phi ,A] &=&-i\int_0^1d\xi \int dtd^Dx\phi (x,t)G(x,t;x^{^{\prime
}},t^{^{\prime }},[\xi \phi ])|_{\stackrel{t^{\prime }\rightarrow t}{%
x^{\prime }\rightarrow x}}  \nonumber \\
&&+{\frac 1{2TL^D}\sum_{q,\Omega }\frac 1{v(q)}\phi (-q,-\Omega )\phi
(q,\Omega )}  \label{3-4}
\end{eqnarray}
The eigen-functional equation of the propagator operator $\widehat{M}(x,t)$
reads, 
\begin{equation}
\left[ i\hbar \frac \partial {\partial t}+\mu -\frac 1{2m}(\hat{p}-g{\bf A)}%
^2+\phi (x,t)\right] \Psi _{k\omega }(x,t,[\phi ])=E_{k\omega }[\phi ]\Psi
_{k\omega }(x,t,[\phi ])  \label{3-5}
\end{equation}
where the eigen-values $E_{k\omega }[\phi ]=\hbar \omega -\varepsilon
(k)+\Sigma _k[\phi ]$, $\varepsilon (k)=(\hbar {\bf k})^2/(2m)-\mu $, and
the self-energy $\Sigma _k[\phi ]=\int_0^1d\xi \int dtd^Dx{\Psi
_k^{*}(x,[\xi \phi ])[\phi (x)+\frac g{2m}{\bf A\cdot }{(2}\widehat{p}{-g%
{\bf A})}]\Psi _k(x,[\xi \phi ])}$ is a regular function, and independent of 
$\omega $ (see below equation \ref{3-6}). Using the orthogonality and
completeness of the eigen-functionals $\Psi _{k\omega }(x,t,[\phi ])$, the
Green's functionals $G(x,t;x^{^{\prime }},t^{^{\prime }},[\phi ])$ can be
represented by these eigen-functionals, 
\begin{equation}
G(x,t;x^{^{\prime }},t^{^{\prime }},[\xi \phi ])=\sum_{k\omega }\frac 1{%
E_{k\omega }[\phi ]}\Psi _{k\omega }(x,t,[\xi \phi ])\Psi _{k\omega
}^{\dagger }(x^{^{\prime }},t^{^{\prime }},[\xi \phi ])  \label{3-5b}
\end{equation}
The eigen-functionals $\Psi _{k\omega }(x,t,[\phi ])$ are the
eigen-functional wave functions of the electrons for a definite HS field $%
\phi (x,t)$, and have the following exact expressions, 
\begin{equation}
\Psi _{k\omega }(x,t,[\xi \phi ])=A_k\left( \frac 1{TL^D}\right)
^{1/2}e^{Q_k(x,t,\xi )}e^{i{\bf k}\cdot {\bf x}-i(\omega +\Sigma _k[\phi ])t}
\label{3-6}
\end{equation}
where $A_k$ is the normalization constant, and $T\rightarrow \infty $ is the
time length of the system (for finite $T$, the $\Psi _{k\omega }(x,t,[\xi
\phi ])$ satisfy the boundary condition $\Psi _{k\omega }(x,t+T,[\xi \phi
])=-\Psi _{k\omega }(x,t,[\xi \phi ])$). These eigen-functionals are
composed of two parts, one represents the free electrons, and another one
represents the correlation of the electrons produced by the electron
interaction. Thus we can formally write 
\begin{equation}
\Psi _{k\omega }(x,t,[\xi \phi ])=\psi _{k\omega }(x,t)A_ke^{Q_k(x,t,\xi
)}e^{-i\Sigma _k[\phi ]t}  \label{3-6b}
\end{equation}
where $\psi _{k\omega }(x,t)$ are the wave-functions of the free electrons.
The phase fields $Q_k(x,t,\xi )$ satisfy the usual Eikonal-type equation
with the condition $Q_k(x,t,\xi )=0$ as $\phi (x,t)=0$ and ${\bf A}(x,t)=0,$ 
\begin{eqnarray}
\left( i\hbar \frac \partial {\partial t}-\frac{\hat{p}^2}{2m}-(\frac \hbar m%
{\bf k}-\frac{\xi g}m{\bf A})\cdot \hat{p}\right) Q_k(x,t,\xi )- &&\frac{[%
\hat{p}Q_k(x,t,\xi )]^2}{2m}  \nonumber \\
-\frac{\xi (g{\bf A})^2}{2m}+\frac{\hbar \xi g}m{\bf k}\cdot {\bf A}+ &&\xi
\phi (x,t)=0  \label{3-7a}
\end{eqnarray}
which can be exactly solved by a series expansion of the HS field $\phi
(x,t) $ and the transverse gauge fields ${\bf A}(x,t)$ and/or by computer
calculations\cite{20,21}. It is noted that the $\phi (x,t)$ and ${\bf A}%
(x,t) $ dependence of the eigen-functionals $\Psi (x,t,[\xi \phi ])$ is
completely determined by the phase fields $Q_k(x,t,\xi )$.

Substituting equation (\ref{3-6}) into equation (\ref{3-5b}), the Green's
functionals can be written as, 
\begin{equation}
G(x,t;x^{^{\prime }},t^{^{\prime }},[\xi \phi ])=\frac i{L^D}\sum_k\theta
(\Sigma _k[\phi ]-\varepsilon (k))|A_k|^2e^{i{\bf k}\cdot ({\bf x-x}^{\prime
})-i\varepsilon (k)(t-t^{\prime })}e^{Q_k(x,t,\xi )+Q_k^{*}(x^{\prime
},t^{\prime },\xi )}  \label{3-7b}
\end{equation}
In order to calculate the Green's functionals $G(x,t;x^{^{\prime
}},t^{^{\prime }},[\xi \phi ])|_{\stackrel{t^{\prime }\rightarrow t}{%
x^{\prime }\rightarrow x}}$, we take the following regular procedure, 
\begin{equation}
G(x,t;x^{^{\prime }},t^{^{\prime }},[\xi \phi ])|_{\stackrel{t^{\prime
}\rightarrow t}{x^{\prime }\rightarrow x}}=\frac 12\lim_{\eta \rightarrow
0^{+}}\left( G(x,t;x-\eta ,t,[\xi \phi ])+G(x,t;x+\eta ,t,[\xi \phi ])\right)
\label{3-7c}
\end{equation}
then we obtain the following exact functional expression of the action, 
\begin{eqnarray}
&&S[\phi ,A]=\frac 12\int dtd^Dx\left[ F(x,x-\eta ,t)+F(x,x+\eta ,t)\right]
_{\eta \rightarrow 0^{+}}  \nonumber \\
&&+{\frac 1{2TL^D}\sum_{q,\Omega }\frac 1{v(q)}\phi (-q,-\Omega )\phi
(q,\Omega )}  \label{3-7d} \\
&&F(x,x^{\prime },t)=T\int_0^1d\xi \sum_k\theta (\Sigma _k[\phi
]-\varepsilon (k))\frac{\phi (x,t)e^{i{\bf k}\cdot ({\bf x-x}^{\prime
})}e^{Q_k(x,t,\xi )+Q_k^{*}(x^{\prime },t,\xi )}}{\int dtd^Dxe^{Q_k(x,t,\xi
)+Q_k^{*}(x,t,\xi )}}  \nonumber
\end{eqnarray}
It becomes very clear that this kind of problems end in to solve the
differential equation (\ref{3-7a}) of the phase fields $Q_k(x,t,\xi )$,
which can be solved by a series expansion of the boson field $\phi (x,t)$
and the transverse gauge fields ${\bf A}(x,t)$. With this action and the
eigen-functionals, we can calculate a variety of correlation functions, such
as the single electron Green's function, by taking the functional average on
the boson field $\phi (x,t)$ and the transverse gauge fields ${\bf A}(x,t)$, 
\begin{eqnarray}
G(x-x^{\prime },t-t^{\prime }) &=&i<G(x,t;x^{^{\prime }},t^{^{\prime
}},[\phi ])>_{\phi ,{\bf A}}  \nonumber \\
&=&i\frac{\int \prod D\phi \prod D{\bf A}G(x,t;x^{^{\prime }},t^{^{\prime
}},[\phi ])e^{iS[\phi ,A]}}{\int \prod D\phi \prod D{\bf A}e^{iS[\phi ,A]}}
\label{3-7e}
\end{eqnarray}
In general, it is difficult to exactly calculate the action, in order to
study the low energy behavior of the system, we can approximately calculate
it. After taking some approximations, we can obtain the following effective
action (omitting constant terms), 
\begin{eqnarray}
S[\phi ,A] &=&{\frac 1{2TL^D}\sum_{q,\Omega }\frac{\phi (-q,-\Omega )\phi
(q,\Omega )}{v(q)}}  \nonumber \\
&&+{\frac 12\int dtd^Dx\phi (x,t)\left[ F_1(x,t,\delta )+F_2(x,t)\right] _{%
{\bf \eta }\rightarrow 0}}  \label{3-8}
\end{eqnarray}
where $F_1(x,t,\eta )=-(1/(2\pi )^D)\int d^Dk\theta (-\varepsilon (k))\sin (%
{\bf k}\cdot {\bf \eta }){\bf \eta }\cdot {\bf \nabla }Q_k^I(x,t)$, $%
F_2(x,t)=(2/(2\pi ^D))\int d^Dk\theta (-\varepsilon (k))Q_k^R(x,t)$. We have
taken $\theta (\Sigma _k[\phi ]-\varepsilon (k))\cong \theta (-\varepsilon
(k))$ and $e^{Q_k(x,t,\xi )+Q_k^{*}(x^{\prime },t,\xi )}\cong 1+Q_k(x,t,\xi
)+Q_k^{*}(x^{\prime },t,\xi ),$ and written the phase fields as $Q_k(x,t,\xi
=1)=Q_k(x,t)=Q_k^R(x,t)+iQ_k^I(x,t)$. It is worthily noted that the
contribution of the ''kinetic'' energy of the electrons to the action is
composed of two parts, one is from the real phase field $Q_k^R(x,t)$, and
another one is from the imaginary phase field $Q_k^I(x,t)$. Due to the Fermi
surface structure of the system, the momentum integral in $F_1(x,t,\eta )$
can be written as\cite{22}, $\int d^Dk=S_{D-1}\int d|k||k|^{D-1}\int_0^\pi
d\theta (\sin \theta )^{D-2}$, where $S_D=2\pi ^{D/2}/\Gamma (D/2)$. As $%
D\geq 2$, the integration of $\sin ({\bf k}\cdot {\bf \eta })$ is regular,
thus as ${\bf \eta }\rightarrow 0$ the function $F_1(x,t,\eta )=0$, the
imaginary phase field $Q_k^I(x,t)$ does not contribute to the action. Only
at $D=1$, it has contribution to the action, where the real phase field $%
Q_k^R(x,t)$ is zero (see below). This property is independent of the
electron interaction, it is completely determined by the Fermi surface
structure of the system.

For a 1D interacting electron gas (taking ${\bf A}(x,t)=0$), the electron
energy spectrum near its Fermi level $\pm k_F$ can be written as, $%
\varepsilon (k)=\pm v_Fk$, where $v_F$ is the Fermi velocity. The branch $%
\varepsilon (k)=v_Fk$ represents the right-moving electrons, and the branch $%
\varepsilon (k)=-v_Fk$ the left-moving electrons. In general, the electron
interaction term reads, $(1/L)\sum_qv(q)\rho _R(q)\rho _L(-q)$, where $\rho
_{R(L)}(q)$ are the right- and left-moving electron densities, respectively,
and $v(q)\cong V$, a constant. To decouple this four-fermion interaction, we
introduce two HS fields $\phi _{R(L)}(x,t)$, and have two sets of
wave-functions of the right- and left-moving electrons for the definite $%
\phi _{R(L)}(x,t)$, respectively, 
\begin{equation}
\Psi _{R(L)k\omega }(x,t,[\xi \phi ])={\left( \frac 1{TL}\right)
^{1/2}e^{Q_{R(L)}(x,t,\xi )}e^{ikx-i(\omega -\Sigma _{R(L)}[\phi ])t}}
\label{3-8b}
\end{equation}
where $\Sigma _{R(L)}[\phi ]$ is a regular quantity, and independent of $k$
and $\omega $. The phase fields $Q_{R(L)}(x,t,\xi )$ are independent of $k$,
and satisfy the simplified Eikonal equation, 
\begin{equation}
{\left( i\frac \partial {\partial t}\pm iv_F\frac \partial {\partial x}%
\right) Q_{R(L)}(x,t,\xi )-\xi \phi _{R(L)}(x,t)}=0  \label{3-8c}
\end{equation}
These linear differential equations can be easily solved, and the phase
fields $Q_{R(L)}(x,t,\xi =1)=Q_{R(L)}(x,t)$ are imaginary because the HS
fields $\phi _{R(L)}(x,t)$ are real. The action (\ref{3-7d}) becomes the
following simple form\cite{23}, 
\begin{equation}
S[\phi ]=\frac 1{TL}\sum_{q,\Omega }\left[ \frac 1{4\pi }\frac q{q-\Omega }%
|\phi _R|^2+\frac 1{4\pi }\frac q{q+\Omega }|\phi _L|^2+\frac 1V\phi
_R(-q,-\Omega )\phi _L(q,\Omega )\right]  \label{3-8d}
\end{equation}
where $\phi _{R(L)}=\phi _{R(L)}(q,\Omega )$. It is worthily noted that the
imaginary phase fields $Q_{R(L)}(x,t)$ not only determine the electron
correlation, but also contribute to the action. This is qualitatively
different from that in 2D and 3D electron gases, where the imaginary part of
the phase field $Q_k(x,t)$ does not contribute to the action due to their
Fermi surface structures.

As $D\geq 2$, for simplicity, we can solve the Eikonal equation by
neglecting the quadratic term $({\bf \nabla }Q_k)^2$. This approximation is
reasonable because for the long-range Coulomb interaction, only the states
near the Fermi surface with momentum $q<q_c$ ($q_c\ll k_F$) are important in
the low energy regime, and for the smooth function $Q_k(x,t),$ this
quadratic term is proportional to $(q_c/k_F)^2\sim 0$. Under this
approximation, we can obtain the effective action, 
\begin{equation}
S_{eff.}=\frac 1{2TL^D}\sum_{q,\Omega }\left[ \left( \frac 1{v(q)}-\chi
(q,\Omega )\right) |\phi (q,\Omega )|^2+\Pi _{ij}(q,\Omega )A_i(-q,-\Omega
)A_j(q,\Omega )\right]  \label{3-9}
\end{equation}
where $\chi (q,\Omega )$ is usual Lindhard function\cite{mahan}, and $\Pi
_{ij}(q,\Omega )=(i\gamma \Omega /q+\chi q^2)(\delta _{ij}-q_iq_j/q^2)$ the
propagator of the transverse gauge fields $A_i(q,\Omega )$ produced by the
electron-hole excitations\cite{lee}. In fact, the above approximation is
equivalent to usual random-phase approximation (RPA). However, at present
framework, it gives more useful and important informations than usual RPA
method, because we have the well-defined phase fields, and can use their
imaginary part to determine the electron correlation.

\bigskip

\begin{center}
\smallskip B. One-band Hubbard model on a square lattice
\end{center}

\bigskip

In the above subsection, we have considered the D-dimensional electron gas
with the transverse gauge fields in the continuous coordinate space, now for
simplicity we consider an one-band Hubbard model on a square lattice, but
this method can be easily extended to other lattice models. The one-band
Hubbard model on a square lattice is presented by the Hamiltonian, 
\begin{equation}
H=-t\sum_{<ij>\sigma }(\widehat{c}_{i\sigma }^{\dagger }\widehat{c}_{j\sigma
}+\widehat{c}_{j\sigma }^{\dagger }\widehat{c}_{i\sigma })-\mu \sum_{i\sigma
}\widehat{c}_{i\sigma }^{\dagger }\widehat{c}_{i\sigma }+U\sum_in_{i\uparrow
}n_{i\downarrow }  \label{3b-1}
\end{equation}
Introducing the Lagrangian multiplier fields $\phi _{i\sigma }(t)$, we have
the following action, 
\begin{equation}
S[\phi ,n]=\sum_{<ij>\sigma }\int dtc_{i\sigma }^{\dagger }(t)M_{ij\sigma
}(t)c_{j\sigma }(t)-\sum_{i\sigma }\int dt[n_{i\sigma }(t)\phi _{i\sigma
}(t)+\frac 12Un_{i\uparrow }(t)n_{i\downarrow }(t)]  \label{3b-2}
\end{equation}
where $M_{ij\sigma }(t)=\delta _{ij}[i\partial _t+\mu +\phi _{i\sigma
}(t)]+t(\gamma _{ij}+\gamma _{ji})$ is a $N\times N$ matrix, where $\gamma
_{ij}=1$ for $j=i\pm 1$, and $\gamma _{ij}=0$ for other $j$. After
integrating out the electron fields $c_{i\sigma }(t)$, we obtain the
following action, 
\begin{equation}
S[\phi ,n]=-iTr\ln \left( M_{ij\sigma }(t)\right) -\sum_{i\sigma }\int
dt[n_{i\sigma }(t)\phi _{i\sigma }(t)+\frac 12Un_{i\uparrow
}(t)n_{i\downarrow }(t)]  \label{3b-3}
\end{equation}
Using the formula, $-iTr\ln \left( M_{ij\sigma }(t)\right) =-iTr\ln \left(
M_{ij}^{(0)}\right) -i\sum_{i\sigma }\int_0^1d\xi \int dt\phi _{i\sigma
}(t)G_\sigma (x_i,t;x_j,t^{\prime },[\xi \phi ])|_{\stackrel{t^{\prime
}\rightarrow t}{x_j\rightarrow x_i}}$, where $M_{ij}^{(0)}=\delta
_{ij}(i\partial _t+\mu )+t(\gamma _{ij}+\gamma _{ji})$, and $M_{ij\sigma
}(t)G_\sigma (x_j,t;x_l,t^{\prime },[\phi ])=\delta _{il}\delta (t-t^{\prime
})$, this action can be re-written as (neglecting the constant term), 
\begin{eqnarray}
S[\phi ,n] &=&-i\sum_{i\sigma }\int_0^1d\xi \int dt\phi _{i\sigma
}(t)G_\sigma (x_i,t;x_j,t^{\prime },[\xi \phi ])|_{\stackrel{t^{\prime
}\rightarrow t}{x_j\rightarrow x_i}}  \nonumber \\
&&-\sum_{i\sigma }\int dt[n_{i\sigma }(t)\phi _{i\sigma }(t)+\frac 12%
Un_{i\uparrow }(t)n_{i\downarrow }(t)]  \label{3b-4}
\end{eqnarray}
The eigen-functional equation of the matrix operator $M_{ij\sigma }(t)$
reads, 
\begin{equation}
M_\sigma (t)\Psi _{k\omega \sigma }(t,[\phi ])=E_{k\omega \sigma }[\phi
]\Psi _{k\omega \sigma }(t,[\phi ])  \label{3b-5}
\end{equation}
where the eigen-values $E_{k\omega \sigma }[\phi ]=\omega -\varepsilon
_k+\Sigma _{k\sigma }[\phi ]$, $\varepsilon _k=-2t[\cos (ak_x)+\cos
(ak_y)]-\mu $, and $\Sigma _{k\sigma }[\phi ]=\int_0^1d\xi \int dt\Psi
_{k\omega \sigma }^{\dagger }(t,[\phi ])\phi _\sigma (t)\Psi _{k\omega
\sigma }(t,[\phi ])$, where $\phi _\sigma (t)=\left( \phi _{i\sigma
}(t)\delta _{ij}\right) $ is a $N\times N$ matrix. The eigen-functionals $%
\Psi _{k\omega \sigma }(t,[\phi ])$ have the following general expression, 
\begin{equation}
\Psi _{k\omega \sigma }(t,[\xi \phi ])=A_k\left( \frac 1{TL^2}\right)
^{1/2}\left( 
\begin{array}{l}
e^{i{\bf k}\cdot {\bf x}_1}e^{Q_{k\sigma }(x_1,t,\xi )} \\ 
\qquad \ \ \vdots  \\ 
e^{i{\bf k}\cdot {\bf x}_N}e^{Q_{k\sigma }(x_N,t,\xi )}
\end{array}
\right) e^{-i(\omega +\Sigma _{k\sigma }[\phi ])t}  \label{3b-6}
\end{equation}
where $A_k$ is a normalization constant, and the phase fields $Q_{k\sigma
}(x_i,t,\xi )$ satisfy the following differential equation, 
\begin{eqnarray}
i\partial _tQ_{k\sigma }(x_i,t,\xi ) &&+\xi \phi _{i\sigma }(t)-\gamma
_k(1-e^{-Q_{k\sigma }(x_i,t,\xi )+Q_{k\sigma }(x_{i+1},t,\xi )})  \nonumber
\\
- &&\gamma _k^{*}(1-e^{-Q_{k\sigma }(x_i,t,\xi )+Q_{k\sigma }(x_{i-1},t,\xi
)})=0  \label{3b-7}
\end{eqnarray}
where $\gamma _k=t(e^{iak_x}+e^{iak_y})$. Using the orthogonality and
completeness of the eigen-functionals $\Psi _{k\omega \sigma }(t,[\xi \phi ])
$, the Green's functionals can be written as, 
\begin{eqnarray}
G_\sigma (x_i,t;x_j,t^{\prime },[\xi \phi ]) &=&\frac i{L^2}\sum_k\theta
(\Sigma _k[\phi ]-\varepsilon (k))|A_k|^2e^{i{\bf k}\cdot ({\bf x}_i{\bf -x}%
_j)-i\varepsilon (k)(t-t^{\prime })}  \nonumber \\
&&\times e^{Q_{k\sigma }(x_i,t,\xi )+Q_{k\sigma }^{*}(x_j,t^{\prime },\xi )}
\label{3b-8}
\end{eqnarray}
then the action of the system reads, 
\begin{eqnarray}
&&S[\phi ,n]=\frac 12\sum_{i\sigma }\int dt\left[ F_\sigma (x_i,x_i-\eta
,t)+F_\sigma (x_i,x_i+\eta ,t)\right] _{\eta \rightarrow 0^{+}}  \nonumber \\
&&-\sum_{i\sigma }\int dt[n_{i\sigma }(t)\phi _{i\sigma }(t)+\frac 12%
Un_{i\uparrow }(t)n_{i\downarrow }(t)]  \label{3b-9} \\
&&F_\sigma (x_i,x_j,t)=T\int_0^1d\xi \sum_k\theta (\Sigma _k[\phi
]-\varepsilon (k))\frac{\phi (x_i,t)e^{i{\bf k}\cdot ({\bf x}_i{\bf -x}%
_j)}e^{Q_{k\sigma }(x_i,t,\xi )+Q_{k\sigma }^{*}(x_j,t,\xi )}}{\sum_i\int
dte^{Q_{k\sigma }(x_i,t,\xi )+Q_{k\sigma }^{*}(x_i,t,\xi )}}  \nonumber
\end{eqnarray}
It is clear that this action is very similar to that of the electron gas in
the continuous coordinate space, the difference between them is that the
phase fields $Q_{k\sigma }(x_i,t,\xi )$ and $Q_k(x,t,\xi )$ satisfy
different differential equations (\ref{3b-7}) and (\ref{3-7a}),
respectively. With equations (\ref{3b-6}), (\ref{3b-7}) and (\ref{3b-9}), we
can completely determine the low energy behavior of the one-band Hubbard
model on the square lattice, at least we can do that by numerical
calculations, because we believe that with the help of these equations we
may need much less computer time than previous methods for the large lattice
site numbers.

Now there is a common consensus that the physical properties of the high Tc
cuprate superconductivity is determined by the 2D strongly correlated
electron gas in their copper-oxide plane(s) which can be approximately
represented by the one-band Hubbard model\cite{7}. If this is true, our
present results can provide enough informations to determine the physical
properties of the high Tc cuprate superconductivity, because using equation (%
\ref{e8}), we can exactly determine their ground state energy and ground
state electron density field, and using equations (\ref{3b-6}) and (\ref
{3b-9}), we can calculate their variety of correlation functions and
response functions to external fields.

\begin{center}
\bigskip

{\bf IV. Ground state wave function of the systems}

\bigskip
\end{center}

\smallskip In principle, we can calculate the ground state wave function of
any systems by using the eigen-functionals. For boson systems, at zero
temperature they have the Bose-Einstein condensation, and the bosons occupy
the state with the momentum $k=0$, and have the eigen-functional wave
function $\Psi _0(x,[\phi ])$. However, for fermion systems, the phase
fields may depend on the momentum $k$ of the fermions, it becomes very
difficult to calculate their ground state wave functions. In general, the
ground state wave function of the fermion system can be written as, 
\begin{equation}
\Psi (x_1,x_2,...,x_N)=<\left\| \Psi _{k_i}(x_j,[\phi ])\right\| >_\phi 
\label{4-0}
\end{equation}
where $k_N=k_F$ is the Fermi momentum, and the functional average over the
HS boson field $\phi (x)$ can be done by taking $Q_{k_i}(x_j)\equiv
Q_{k_i}(x_j,t)$ and $\phi (x_i)\equiv \phi (x_i,t)$. Here we only consider
the boson systems and the two-dimensional electron gas under an external
magnetic field there appears the fractional quantum Hall effect for enough
strong magnetic field.

\medskip

\begin{center}
\smallskip A. Boson systems
\end{center}

\medskip

For boson systems, due to the condensation of the bosons, we have the
following ground state wave-function, 
\begin{eqnarray}
&&\Psi (x_1,x_2,...,x_N)={\ \left( \frac{A_0}{L^{D/2}}\right) ^N<\Psi
(x_1,x_2,...,x_N,[\phi ])>_\phi }  \nonumber \\
&&\Psi (x_1,x_2,...,x_N,[\phi ])={\ e^{\sum_{i=1}^NQ_0(x_i,\xi =1)}}
\label{4-1}
\end{eqnarray}
where $<...>_\phi $ means the functional average over the boson field $\phi
(x)$. The functional $\Psi (x_1,x_2,...,x_N,[\phi ])$ is very similar to the
generalized London wave-function\cite{17}, where $f(x_i)=\exp
\{Q_0^R(x_i,\xi =1)\}$ and $S(x_i)=Q_0^I(x_i,\xi =1)$. To calculate this
functional average, we need knowing the action of the systems. Without
external fields and constraints, the phase fields $Q_0(x,t,\xi )$ satisfy
the simple Eikonal equation with the condition $Q_0(x,t,\xi )=0$ as $\phi
(x,t)=0$, 
\begin{equation}
\left( i\hbar \frac \partial {\partial t}-\frac{\hat{p}^2}{2m}\right)
Q_0(x,t,\xi )-\frac{[\hat{p}Q_0(x,t,\xi )]^2}{2m}+\xi \phi (x,t)=0
\label{4-2}
\end{equation}
If we only keep the linear terms, we have the following simple solution, 
\begin{equation}
Q_0(x,t,\xi )=-\frac \xi {TL^D}\sum_{q,\Omega }\frac{\phi (q,\Omega )}{\hbar
\Omega -\epsilon _q}e^{i{\bf q\cdot x}-i\Omega t}  \label{4-3}
\end{equation}
where $\epsilon _q=\hbar ^2q^2/(2m)$. Substituting it into (\ref{3-8}), we
obtain the following effective action with $k=0$, 
\begin{equation}
S_0[\phi ]={\frac 1{TL^D}\sum_{q,\Omega }}\left( -\frac{N\epsilon _q}{(\hbar
\Omega )^2-\epsilon _q^2}+\frac 1{2v(q)}\right) \phi (-q,-\Omega )\phi
(q,\Omega )  \label{4-4}
\end{equation}
It is worthily noted that with this action we cannot obtain the correct
excitation spectrum of the systems, because it is obtained by using the
phase fields $Q_0(x,t,\xi )$. In order to get the correct excitation
spectrum of the bosons, we need knowing the action obtained by the phase
fields $Q_k(x,t,\xi )$. However, we can approximately use the action $%
S_0[\phi ]$ to calculate the functional average over the boson field $\phi
(x,t)$ in equation (\ref{4-1}) by taking $Q_0(x_i,\xi =1)\equiv
Q_0(x_i,t,\xi =1)$. The detail expression form of the ground state wave
function depends on the interaction potential $v(q)$, but it can be
generally written as, 
\begin{equation}
\Psi (x_1,x_2,...,x_N)\sim e^{-\sum_{i<j}U(x_i-x_j)}  \label{4-5}
\end{equation}
where $U(x_i-x_j)$ is uniquely determined by the interaction potential $%
v(x_i-x_j)$. It is clearly seen that this ground state wave-function is
uniquely determined by single effective potential function, and has the
expression very similar to usual correlated basis functions\cite{16} that
are the type of wave-function most often employed in the study of the ground
state properties of $^4He$.

\medskip

\begin{center}
\smallskip B. Two-dimensional electron gas under an external magnetic field
\end{center}

\medskip

The ground state wave function of this system cannot be written as a simple
form even if for the lowest Landau level ($n=0$), because the phase fields $%
Q_{nl}(x,\xi =1)$ depend on the quantum numbers $n$ and $l$. In the lowest
Landau level, if we approximately take $Q_{0l}(x,\xi =1)\sim Q_{00}(x)$, we
can obtain the following expression of the ground state wave-function, 
\begin{equation}
\Psi
(x_1,x_2,...,x_N)=A\prod_{i=1}^Nz_i\prod_{i>j}(z_i-z_j)e^{%
\sum_{i=1}^N|z_i|^2}<e^{\sum_{i=1}^NQ_{00}(x_i)}>_\phi  \label{4-6}
\end{equation}
where $A$ is a normalization constant, and the factor $\prod_{i>j}(z_i-z_j)$
guarantees the anti-commutation of the electrons. The last factor $<\exp
\{\sum_{i=1}^NQ_{00}(x_i)\}>_\phi $ is very similar to the ground state
wave-function of the boson systems (\ref{4-5}), and it is very clear that
this factor is the contribution of the electron interactions. Using the
procedure as above for the boson systems, we can obtain an approximate
action by only keeping the linear terms in solving the differential equation
of the phase field $Q_{00}(x)$, then with this action we can calculate the
functional average in (\ref{4-6}). Only for long range Coulomb interaction $%
v(q)=e^2/(4\pi q^2)$ can one have $<\exp \{\sum_{i=1}^NQ_{00}(x_i)\}>_\phi
\sim \prod_{i>j}|z_i-z_j|^\gamma $, where $\gamma $ is a dimensionless
constant, but at present approximation, it is hard to determine the relation
between $\gamma $ and the filling factor of the lowest Landau level. Thus
this ground state wave function is different from the Laughlin's trial
wave-functions of the odd-denominator fractional quantum Hall states.
However, we believe that with the equations of the ground state energy and
the ground state wave function (\ref{17}) and (\ref{4-6}), we can obtain
more important informations of the fractional quantum Hall effects than that
by the Laughlin's trial wave-functions, because they are directly derived
from the microscopic theory.

\begin{center}
\bigskip

{\bf V. Unified description of strongly and weakly correlated electron gases}

\bigskip
\end{center}

In this section, we show that the strongly and weakly correlated electron
gases can be unifiably represented under present theoretical framework. In
fact, as shown in the section II and /or section III, the problems of the
quantum many-particle systems end in to solve the differential equation of
the phase fields $Q_k(x,t,\xi )$. Thus the phase fields $Q_k(x,t,\xi )$ are
the key parameters hidden in the quantum many-particle systems, and
completely determine their low energy behavior. It is natural that we can
unifiably represent the quantum many-particle systems with the help of the
phase fields $Q_k(x,t,\xi )$.

It is simple to prove that the 1D interacting electron gas is a strongly
correlated system even for very weak electron interaction $V\sim 0$. With
the action (\ref{3-8d}), by simple calculation we can obtain the relations, 
\begin{eqnarray}
&<&\psi _{R(L)k\omega }(x,t)\Psi _{R(L)k\omega }^{\dagger }(x^{\prime
},t,[\phi ])>_\phi \sim {\left( \frac 1L\right) ^\alpha e^{ik(x-x^{\prime })}%
}  \nonumber \\
&<&e^{Q_{R(L)}(x,t)}e^{-Q_{R(L)}(x^{\prime },t)}>_\phi \sim e^{-2\alpha \ln
|x-x^{\prime }|}{,\;\;\;|x-x^{\prime }|\rightarrow \infty }  \label{5-1}
\end{eqnarray}
where $\alpha \sim (1-V/(2\pi \hbar v_F))/2$ for $V\sim 0$, is the
dimensionless coupling strength parameter, the $\psi _{R(L)k\omega }(x,t)$
are the wave-functions of the right(left)-moving free electrons, and $%
<\cdots >_\phi $ means taking functional average over the HS fields $\phi
_{R(L)}(x,t)$. The first equation presents the zero overlap between the
eigen-functional wave functions $\Psi _{R(L)k\omega }(x,t,[\phi ])$ of the
interaction electrons and the wave functions of the free electrons ($V=0$)
as $L\rightarrow \infty $, thus the states of the interacting electron gas
does not have one-to-one correspondence via adiabatic continuation with
those of the free electron gas. The second equation presents the strong
electron correlation, in which the electron correlation length is infinity
even for weak electron interaction. Due to this strong electron correlation,
the low energy excitation modes of the 1D interacting electron gas are those
collective excitation modes, such as the charge and spin density waves, and
there are not well-defined quasi-particles (holes) near the two Fermi levels 
$\pm k_F$. Thus the Tomonaga-Luttinger liquid theory is a strongly
correlated theory, and is universal for 1D interacting electron gases.

For a 3D electron gas with long-range Coulomb interaction $v(q)=e^2/(4\pi
q^2)$, by simple calculation, we have the relations, 
\begin{eqnarray}
&&{<\psi _{k\omega }(x,t)\Psi _{k\omega }^{\dagger }(x^{\prime },t,[\phi
])>_\phi }\sim {Z_ke^{i{\bf k}\cdot ({\bf x}-{\bf x^{\prime }})}}  \nonumber
\\
&&{<e^{iQ_k^I(x,t)}e^{-iQ_k^I(x^{\prime },t)}>_\phi }\sim {%
e^{z_k^I(x-x^{\prime })}}  \label{5-3}
\end{eqnarray}
where $Z_k$ is finite, and $z_k^I(x)$ is a smooth function. As $|{\bf x}%
|\rightarrow \infty $, the function $z_k^I(x)$ goes to zero. The first
equation presents that the eigen-functionals $\Psi _{k\omega }(x,t,[\phi ])$
have large overlap with the eigen-functions $\psi _{k\omega }(x,t)$, and the
second equation presents that there is only weak electron correlation even
for long-range Coulomb interaction. Thus the fundamental assumption of the
Landau Fermi liquid theory is satisfied, and the Landau Fermi liquid theory
is universal for 3D interacting electron gases.

For a 2D electron gas with the long-range Coulomb interaction $%
v(q)=e^2/(4\pi q^2)$, the situation is different from that in the 3D
electron gas. By simple calculation, we can obtain the relations, 
\begin{eqnarray}
&&{<\psi _{k\omega }(x,t)\Psi _{k\omega }^{\dagger }(x^{\prime },t,[\phi
])>_\phi }\sim {\left( \frac 1{q_cL}\right) ^\beta e^{i{\bf k}\cdot ({\bf x}-%
{\bf x^{\prime }})}}  \nonumber \\
&&{<e^{iQ_k^I(x,t)}e^{-iQ_k^I(x^{\prime },t)}>_\phi }\sim {e^{-2\beta \ln
(q_c|{\bf x}-{\bf x^{\prime }}|)},\;\;\;|{\bf x}-{\bf x^{\prime }}%
|\rightarrow \infty }  \label{5-4}
\end{eqnarray}
where $\beta =e^2/(2(4\pi )^2\omega _p)$ is the dimensionless coupling
strength parameter, and $\omega _p$ is the plasma frequency. In fact, $%
q_cL\rightarrow \infty $, the first equation presents that the
eigen-functionals $\Psi _{k\omega }(x,t,[\phi ])$ has no (or infinitesimal)
overlap with the eigen-functions $\psi _{k\omega }(x,t)$, and the second
equation presents the strong electron correlation. In this case, the
fundamental assumption of the Landau Fermi liquid theory fails, and the
system shows the non-Landau-Fermi liquid behavior. In the low energy limit,
as $\Omega >q\rightarrow 0$, we have $\chi (q,\Omega )\thicksim q^2/\Omega
^2 $, then we obtain the propagator of the HS field $\phi (x,t)$ by the
action (\ref{3-9}), $G_\phi (q,\Omega )=v(q)/(1-\omega _p^2/\Omega ^2)$.
Using this propagator to calculate the functional average over $\phi (x,t)$,
we obtain the results in (\ref{5-4}). These results are consistent with that
of Ref.\cite{wen} obtained by other method. The higher order terms neglected
in solving equation (\ref{3-7a}) cannot alter these results because in low
energy limit they are proportional to $(q_c/k_F)^{2+n}\rightarrow
0,n=0,1,2,...,$ and can be safely neglected. It is worthy noted that for 2D
and 3D electron gases with long-range Coulomb interactions we have the same $%
G_\phi (q,\Omega )$ in which the screening effects are denoted by the factor 
$1/(1-\omega _p^2/\Omega ^2)$. However, the reason they show different low
energy behavior is that the integral $\int d^dqe^{i{\bf q\cdot x}}/q^2$
appearing in the functional average of equations (\ref{5-3}) and (\ref{5-4})
is divergent (log-type) for $d=2$, while it is convergent for $d=3$.

Even if the asymptotic behavior of the single electron Green's function is
similar to that of the 1D interacting electron gas (see, (\ref{5-1}) and (%
\ref{5-4})), they originate from different physics. However, this difference
can also be clearly seen from their effective actions of the density field.
For 1D interacting electron gases, this action can be easily obtained by
usual bosonization method\cite{13,14}, where the density field has
well-defined propagator (density wave). For the 2D electron gas with
long-range Coulomb interaction, the effective action (\ref{3-9}) can be
written as with the density field (taking ${\bf A}(x,t)=0$), 
\begin{equation}
S_{eff.}[\rho ]=\frac 1{2TL^2}\sum_{q,\Omega }\left( \frac 1{\chi (q,\Omega )%
}-v(q)\right) |\rho (q,\Omega )|^2  \label{5-5}
\end{equation}
There appears a gap in the excitation spectrum of the density field (plasmon
excitations). It was shown\cite{ng} that as $e^2\gg 4\pi $ the system can be
described as a Landau Fermi liquid formed by chargeless quasi-particles
which has vanishing wavefunction overlap with the bare electrons in the
system. For short range and/or weak electron interaction, the 2D interacting
electron gases would show the Landau Fermi liquid behavior in the low energy
region.

Based on the above calculations, it becomes clear that if the non-Fermi
liquid behavior of the electron gases is produced by the (long range)
Coulomb interaction of the electrons, the correlation function of the
imaginary phase fields $Q_k^I(x,t)$ must be a log-type. However, there
exists another type correlation function of the imaginary phase fields $%
Q_k^I(x,t)$, which also induces the non-Fermi liquid behavior of the
electron gases. Here we only consider a simple example of 2D electron gas
with transverse gauge fields ${\bf A}(x,t)$. Solving the differential
equation of the phase fields $Q_k(x,t)$ (\ref{3-7a}) where only keeps the
linear terms, and using the effective action (\ref{3-9}), we can obtain
single electron Green's function, 
\begin{eqnarray}
G(x-x^{\prime },t-t^{\prime }) &=&i<G(x,t;x^{\prime },t^{\prime },[\phi ])>_{%
{\bf A}}  \nonumber \\
&=&-{\ \frac 1{L^2}\sum_k\theta (-\varepsilon (k))e^{i{\bf k}\cdot ({\bf x}-%
{\bf x}^{^{\prime }})-i\epsilon (k)(t-t^{\prime })}e^{P_k^I(x-x^{\prime
},t-t^{\prime })}}  \label{5-6}
\end{eqnarray}
where $P_k^I(x,t)$ comes from the contribution of the imaginary phase fields 
$Q_k^I(x,t)$, and the contribution from the real phase fields $Q_k^R(x,t)$
can be neglected. Near the Fermi surface $k\sim k_F$, we have the relation, 
\begin{equation}
P_{k_F}^I(0,t)\simeq -i^{1/3}ag^2t^{1/3}  \label{5-7}
\end{equation}
where $a=\frac{k_F}{4\pi ^2m}(\chi /\gamma )^{1/3}\int dx(1-\cos
(x))x^{-4/3} $. This result is basic same as that of Ref.\cite{27,28}. Due
to the power-law time dependence of the phase factor $P_{k_F}^I(0,t)$, the
single electron Green's function shows a singular low energy dependence
which violates the quasi-particle excitation assumption of the Landau Fermi
liquid. Thus, the system shows non-Fermi liquid behavior. In this case, the
correlation function of the imaginary phase fields $Q_k^I(x,t)$ has the
power-law form $t^\alpha $, $\alpha <1$, near the Fermi surface. For a 2D
electron gas with long range Coulomb interaction and transverse gauge
fields, the correlation function of the imaginary phase fields $Q_k^I(x,t)$
is composed of two parts, one is a log-type which comes from the
contribution of the HS boson field $\phi (x,t)$, and another one has the
power-law form $t^\alpha $, $\alpha <1$, near the Fermi surface, which is
from the contribution of the transverse gauge fields ${\bf A}(x,t)$. Thus
this system shows the non-Fermi liquid behavior in the low energy region.

\begin{center}
\bigskip

{\bf VI. Discussion and conclusion}

\bigskip
\end{center}

Applying previous perturbation methods to study strongly correlated systems,
we may meet many serious problems, because in these systems due to the
strong correlation among fermions/bosons, there does not exist a small
quantity to be used as a perturbation expansion parameter. The most
prominent character of the strongly correlated systems is the strong
fermion/boson correlation, which completely controls their low energy
behavior. This strong fermion/boson correlation is produced by the
fermion/boson interactions, thus the key point of this kind problems is how
to exactly or accurately treat the particle interaction (potential) terms.
However, the traditional treatment of these systems is starting from a bare
particle Green's function to perturbatively treat the particle interaction
(potential) terms, i.e., exactly treating the kinetic energy term and
perturbatively treating the potential terms. Explicitly, this treatment is
unsuccessful for the strongly correlated systems.

It is natural that one uses another way to treat the strongly correlated
systems, i.e., exactly treating the interaction (potential) terms and
exactly or perturbatively treating the kinetic energy term of the particles,
such as usual bosonization method in treating 1D interacting fermion
systems. Only in this way can one effectively treat the strong particle
correlation. However, to realize this idea, we may meet two problems, one is
how to decouple four-particle interaction (potential) term, and another one
is how to represent the particle correlation strength. The former one can be
done by introducing Lagrangian multiplier or Hubbard-Stratonovich boson
fields, and the latter one can be naturally done by introducing phase fields
in solving the eigen-functional equation of the propagator operator of the
particles. It must be mentioned that after introducing the Lagrangian
multiplier or Hubbard-Stratonovich boson fields, the Hilbert space of the
system is enlarged, and the unphysical particle density field also appears
in the Hamiltonian and/or action. To delete these unphysical states or leave
the original Hilbert space intact, we can add a constraint condition to the
functional ground state energy, such as $\delta E[\phi ,\rho ]/\delta \phi
(x)=0$, or take the functional average over the Lagrangian multiplier $\phi
(x,t)$.

Based on the above idea, we have established an unified theory of the
quantum many-particle systems, which is valid not only for weakly correlated
fermion/boson systems, but also for strongly correlated fermion/boson
systems. This unified theory of the quantum many-particle systems has three
prominent characters, (1). it is completely founded on the basic principles
of quantum mechanics. (2). it has well-defined phase fields used to
represent the particle correlation strength. (3). it has an explicit and
simple exact functional expression of the ground state energy, the action
and the eigen-functional wave functions of the particles, and one can easily
use them to effectively (or exactly) treat general quantum many-particle
problems, and can obtain any accurate result as one hoped by taking high
order corrections. In this unified theory, we have shown that the problems
of the quantum many-particle systems end in solving the differential
equation of the phase fields $Q_k(x,\xi )$ and/or $Q_k(x,t,\xi )$, and given
the exact expression of the ground state energy and action, and the
eigen-functional wave functions of the particles as the functional of the HS
boson fields (and/or transverse gauge fields) and the particle density
fields. With them, we can calculate a variety of correlation functions of
the systems. It is surprising but natural that the physical properties of
the many-particle systems is completely controlled by the differential
equation of the phase fields $Q_k(x,t,\xi )$. The above results have been
easily applied to the systems under external fields and/or constraint
conditions, such as trapped boson systems and 2D electron gas under the
external magnetic field, and to the lattice models, such as the Heisenberg
model and the one-band Hubbard model on the square lattice.

The key points of this unified theory are that: (1). we introduce the
Lagrangian multiplier (or HS) field $\phi (x,t)$ which takes the particle
density $\rho (x,t)=\psi ^{\dagger }(x,t)\psi (x,t)$ as a constraint
condition, so that the four-particle interaction term is decoupled and the
action can only have the quadratic form of the particle (fermion/boson)
fields. (2). by introducing the phase field $Q_k(x,t,\xi )$, which is a
functional of the boson field $\phi (x,t)$ and determined by usual
Eikonal-type equation, we can use it to completely represent the ''kinetic''
energy of the systems. Thus the problems of the quantum many-particle
systems end in to solve the differential equation of the phase fields $%
Q_k(x,t,\xi )$. (3). the phase field $Q_k(x,t,\xi )$ is a key parameter
hidden in the quantum many-particle systems, its imaginary part represents
the particle correlation strength, and its real part only contributes to the
ground state energy and action. (4). we are able to use the particle density
field $\rho (x)$ and $\rho (x,t)$ to exactly represent the ground state
energy and action of the quantum many-particle systems, respectively.

\bigskip

\begin{center}
{\bf Acknowledgment}

\medskip
\end{center}

We appreciate useful discussions with Z. B. Su, T. Xiang, X. Q. Wang, X. Dai
and Z. Y. Weng.

\end{document}